\documentclass{aa}
\usepackage[dvips]{graphicx}
\usepackage{epsfig}

\newcommand{\vmi}{V$-$I}
\newcommand{\bmv}{B$-$V}

\newcommand{\vmo}{(V$-$I)$_0$}

\newcommand{\teff}{T$_{\rm eff}$}

\newcommand{\nli}{$\log$~n(Li)}
\newcommand{\efe}{$\log \epsilon$ (Fe)}

\begin{document}

\title{Membership, lithium, and metallicity in the young open
clusters \object{IC~2602} and \object{IC~2391}:
enlarging the sample  \thanks{Based on observations
carried out at the European Southern Observatory, La Silla, Chile}
\thanks{Tables 2--5 are also available in electronic form
at the CDS via anonymous ftp to cdsarc.u-strasbg.fr (130.79.128.5)
or via http://cdsweb.u-strasbg.fr/cgi-bin/qcat?}}

\author{S. Randich \inst{1}, R. Pallavicini \inst{2}, 
G. Meola \inst{2}, J.R. Stauffer\inst{3}, and Suchitra C. Balachandran 
\inst{4}
}
\offprints{S. Randich, \email{randich@arcetri.astro.it}}
\institute
{Osservatorio Astrofisico di Arcetri, Largo Fermi 5, I-50125
Firenze, Italy\\e-mail: randich@arcetri.astro.it
\and Osservatorio Astronomico di Palermo, Piazza del Parlamento 1, 
I-90134
Palermo, Italy
\and Center for Astrophysics, 60 Garden St., Cambridge, MA 02138, USA
\and Department of Astronomy, University of Maryland, College Park, MD 20740,
USA
}

\date{Received date; accepted date}
\titlerunning{Lithium in IC~2602 and IC~2391}
\authorrunning{S. Randich et al.}

\abstract{
We present lithium abundances for $\sim$50 X-ray selected candidate members of 
the 30--50~Myr old open clusters IC~2602 and IC~2391.
These data enlarge and extend to cooler temperatures previous Li surveys 
of these clusters by Stauffer et al. (\cite{sta89}) and Randich et al. 
(\cite{R97}). 
We also give for the first time an estimate of the metallicity 
of the two clusters which turns out to be close to solar.
Radial velocity measurements together with H$\alpha$ chromospheric emission 
and the presence/absence of other spectroscopic features
are used to ascertain the membership
status for the sample stars not yet confirmed as
cluster members; 
rotational velocities have also been determined for all sample stars. 
Stars more massive than $\sim$ 1 M$_{\odot}$ 
in both clusters show no sign of significant Li depletion, 
while lower mass stars
are all lithium depleted, with the amount of Li depletion
increasing to cooler temperatures. 
We confirm that the late G and early K stars 
in IC~2602 present a star-to-star scatter in Li abundances similar to, but 
not as large as the one in the \object{Pleiades}. 
A scatter is also seen among late--K
and M dwarfs. Unlike in the Pleiades and \object{Alpha
Per} clusters, the scatter among early--K stars in IC~2602 shows 
only marginal correlation with rotation. Our data suggest that
the drop-off of lithium towards 
lower masses may start at an earlier color in
IC~2391 than in IC~2602, but larger cluster samples are needed to confirm
this result. In addition, whereas 
G and early K stars in the two clusters
are, on average, more Li rich than their counterparts in the Pleiades,
a fraction of the coolest stars, in particular in IC~2391, are 
as depleted as the lowest-Li Pleiades stars of the same mass.
If they continue depleting Li on their way to the
main sequence, they are expected 
to be more Li depleted than the Pleiades at the
age of the latter cluster. 
        \keywords{open clusters and associations: 
         individual: IC 2602 -- open clusters
          and associations: individual: IC 2391 -- stars: 
         abundances -- stars: interiors}}
\maketitle
\section{Introduction}
Several papers in the last decade have addressed the issue of the
evolution of lithium abundance in solar--type and lower mass stars
as traced
by open clusters of different ages and metallicities. We refer
to the most recent reviews on this topic for a detailed discussion
of the current status of our understanding of the so-called ``Lithium problem"
(e.g., Deliyannis \cite{del00}; Jeffries \cite{jef00}; Pasquini 
\cite{pas00}) and
briefly summarize here the main results
and open questions raised by lithium surveys of open clusters.
It is now well established that ``standard" models, i.e., models
including convection only as mixing mechanism, are unable to explain
the complex picture emerging from the observational data. 
Extra--mixing processes
and/or mechanisms that inhibit mixing and lithium destruction
are indeed required. Standard models predict that surface lithium 
abundances should depend only on mass, age, and 
on chemical composition, and
that most of the Li depletion in solar--type 
and lower mass stars should occur during pre-main sequence (PMS)
evolution. The observations, in particular the comparison
between the 120 Myr old Pleiades and the 600--800 Myr old \object{Hyades} 
and between the
latter cluster and the 4.5 Gyr old \object{M~67}, show instead that Li depletion
in solar--type stars does take place on the main-sequence (MS) and
that stars down to about 1~M$_{\odot}$ 
reach the Zero Age Main Sequence (ZAMS) retaining their initial
Li content (e.g., Mart\'\i n \& Montes \cite{mm97}; Randich et al. \cite{R97}). 
A star--to--star scatter in Li abundance is observed both among 
otherwise similar solar--type
stars in the old cluster M~67 (e.g., Jones et al. \cite{jon99} and references
therein) and among lower mass stars in young clusters (e.g., Soderblom et al.
\cite{sod93}). Finally, at least for solar--type stars,
metallicity does not seem to significantly affect 
lithium depletion 
(Jeffries et al. \cite{jef98}; Jeffries \& James \cite{jj99}). 
Extra--mixing or, more in
general, non standard processes, seem to be at work not only in solar--type
stars on the MS, but also during the PMS and early MS
evolution of K--type stars. Understanding these 
non--standard processes remains a challenging task,
in spite of the continuously growing body of observational data.

We present here lithium observations of late--type stars in the young clusters
IC~2602 and IC~2391 which have estimated ages ($\sim 30 - 55$ Myr;
see below) intermediate between PMS stars and the Pleiades.
We will mainly focus on low and very low mass stars
in the two clusters, with the purpose of further investigating the issue
of Li depletion in the PMS phase and of the developing of a
scatter in Li abundances among low mass stars.

Before X--ray surveys of these clusters by $ROSAT$,
very little information was available on their memberships. 
$ROSAT$ observations,
besides allowing a study of the X--ray properties of both clusters, 
resulted in the detection of several possible low mass cluster members
(Randich et al. \cite{R95}; Patten \& Simon \cite{ps96}). Optical photometric
and spectroscopic follow-up studies have allowed investigators
to ascertain the membership
status for virtually all the X--ray selected candidates of IC~2391
and for several candidates of IC~2602. Membership was confirmed 
for a large fraction of them, providing a first sample of stars 
to study the evolution of lithium, rotation, and activity between 
the PMS and the Pleiades age (see Prosser et al. \cite{pro96}; 
Stauffer et al. \cite{sta97}
--hereafter S97; Randich et al. \cite{R97} --hereafter R97).
This sample will be further exploited and enlarged
in the present paper. With regards to age,
very recently an age of 55~Myr has been estimated for IC~2391
using the Li depletion boundary method
(Barrado y Navascu\'es et al. \cite{bar99}), 
while color--magnitude (C-M) diagram fittings
for both IC~2391 and IC~2602 give an age of $\sim 30-35$~Myr.
No metallicity determinations for the two clusters were available up to now.
\begin{table*}
\caption{The sample}
 \begin{tabular}{lccclccl}\\ \hline
 & & & & & & & \\
\multispan{8}{IC~2602: sample observed at ESO (April `95) \hfill}\\
 & & & & & & & \\
star 
& V & B$-$V & (V$-$I)$_c$ & H$\alpha$ & v$\sin i$ & v$_{\rm rad}$ & remarks\\ 
(1)  &   \multicolumn{3}{c}{(2)}& (\AA)     & (km/s)      & (km/s) & \\
R1    & 11.57 & 0.91 & 0.91 & $-0.3$ & $\leq 10$ & 16   & member\\
R8A   & 10.41 & 0.65 & 0.66 & $-1.0$ &      26   & 28   & SB;
member?/see text   \\
R24A  & 14.60 & 1.43 & 1.86 & $+$3.0 &      35   & 17   & member \\
R28A  & 15.68 & 2.07 & 2.53 & $+$0.5 & $\leq 10$ & 97   & non member/see text \\
R30   & 15.73 & 1.51 & 2.30 & $+$4.8 &      12   & 12   & member: \\
R31   & 15.08 & 1.59 & 2.24 & $+$4.0 &      42   & 16   & member \\
R37A  & 14.69 & 1.86 & 2.00 & ---    &     120:  & ---  & ?  \\
R38   & 15.72 & 1.53 & 2.50 & $+$4.9 &      60   & 14:  & member \\
R39   & 15.99 & 1.57 & 2.55 & $+$4.1 &      22   & 11   & member: \\
R42C  & 11.56 & 0.80 & 0.87 & abs.   & $\leq 10$ &$-40$ & non member  \\
R46   & 10.70 & 0.67 & 0.82 & $-1.1$ &       18  & 74   & non member  \\
R52   & 12.19 & 1.07 & 1.11 & $-0.9$ &      110: & 12:  & member \\
R54A  & 12.13 & 1.15 & 1.37 & $+$0.2  &      28  & 14   & member \\
R73A  & 16.02 & ---  & 2.49 & $+$3.5  &   30/90? & 20:  & member?/see text \\
 & & & & & &  &\\  \hline 
\end{tabular}
\end{table*}
\begin{table*}
\begin{tabular}{lcccllcl}\\ \hline
 & & & & & &  & \\
\multispan{8}{IC~2602: sample observed at CTIO (January `94) \hfill}\\
 & & & & & & &  \\
star & V & B$-$V & (V$-$I)$_c$ & H$\alpha$ & v$\sin i$ & v$_{\rm rad}$ 
&  remarks\\ 
(1)  &   \multicolumn{3}{c}{(2)} & (\AA) & (km/s) & (km/s) &\\
W79   & 11.57 & 0.83 & 0.85 &  $-$0.9 & 8.0    & 17.3 & \\
R15   & 11.75 & 0.93 & 1.06 &  +0.2   & 7:     & 17.4 & \\  
R24A  & 14.61 & 1.43 & 1.86 &  +1.9   & 34.    & 18.7 & \\
R26   & 15.14 & 1.54 & 2.15 &  +1.0   &  $<$ 6 & 17.7 & \\
R27   & 14.35 & 1.50 & 1.80 &  +1.2   & 10.0   & 17.0 & \\
R31   & 15.08 & 1.59 & 2.24 &  +3.9   & 35.    & 16.3 & \\
R32   & 15.06 & 1.63 & 2.16 &  +1.9   & 9.0    & 18.1 & \\
R38   & 15.72 & 1.53 & 2.50 &  +11.3  & 48.    & 17.3 & \\
R44   & 14.88 & 1.55 & 2.03 &  +1.4   & 6:     & 17.2 & \\
R50   & 14.75 & 1.56 & 2.08 &  +1.4   & 6:     & 16.7 & \\
R53B  & 15.39 & 1.61 & 2.49 &  +5.1   & 100.   & 17.1 & \\
R57   & 15.59 & 1.60 & 2.44 &  +3.2   & $<$ 7  & 19.1 & \\
R66   & 11.07 & 0.68 & 0.83 &  $-$1.0 & 12.0   & 17.3 & \\
R67   & 14.97 & 1.57 & 2.16 &  +1.5   & $<$ 6  & 16.4 & \\
R70   & 10.92 & 0.69 & 0.71 & $-$1.2  & 11.0   & 16.7 & \\
R77   & 14.12 & 1.47 & 1.72 &  +0.4   & $<$ 7  & 17.6 & \\
R82   & 14.98 & 1.62 & 2.51 &  +3.4   & $<$ 7  & 14.6 & SB3\\
R93   & 13.79 & 1.37 & 1.62 &  +0.6   & 8.5    & 18.5 &  \\
R96   & 12.94 & 1.25 & 1.37 & +0.6    & 17.0   & 15.5 &  \\
 & & & & &  & & \\ \hline 
 & & & & &  & & \\ 
\multispan{8}{(1) Names from R97 and Whiteoak (\cite{whi61})\hfill}\\
\multispan{8}{(2) Photometry comes from Prosser et al. (\cite{pro96})\hfill}\\
\end{tabular}
\end{table*}
\begin{table*}
\begin{tabular}{lcccllcl}\\ \hline
 & & & & & & &  \\
\multispan{8}{IC~2391: sample observed at CTIO (January `94) \hfill}\\
 & & & & & & &  \\
star & V & B$-$V & (V$-$I)$_c$ & H$\alpha$ & v$\sin i$ & V$_{\rm rad}$ &
remarks\\ 
\multicolumn{4}{c}{(1)}& (\AA) & (km/s) & (km/s) & \\
VXR3A  & 10.95 & ---    & 0.74  & $-1$1.1 &  10.     & 14.9 &   \\
VXR6A  & 13.77 & ---    & 1.94  & +1.4    &  $<$ 6, $<$ 7 &$-20.6$,49.9 & SB2\\
VXR7   & 9.63  & 0.46   & ---   & $-$1.5  &  21      & 15.6 &   \\
VXR14  & 10.45 & (0.57) & 0.69  & $-1.1$  &  47      & 14. &   \\
VXR16A & 11.84 & ---    & 0.94  & $\sim$0.&  22      & 15.5 &  \\
VXR18A & 13.54 & ---    & 1.53  & $=$0.4  &  8       & 14.6  \\
VXR30  & 9.88  & 0.46   & ---   & $-1$1.4 &  43      & 30.1 &  SB1?\\
VXR31  & 11.22 & ---    & 0.73  & $-1$1.1 &  17      & 22.3 & SB1?\\
VXR44  & 9.69  & 0.42   & ---   & $-$1.7  &  67:     & 14.2 & \\
VXR49B & 14.34 & ---    & 1.89  & +1.0    & 12.2      & 13.7 & \\
VXR50A & 12.54 & ---    & 0.91  &$-$0.15  &  64      & 17. & \\
VXR62A & 11.73 & (0.86) & 0.99  & $\sim$0.&  52      & 12: & \\
VXR65  & 14.13 & ---    & 1.79  & +0.8    &   8      & 14.5 & \\
VXR67A & 11.71 & ---    & 1.03  & $-$0.1  &   8:      & 14.9 & \\
VXR69A & 11.67 & ---    & 0.90  & $-$0.3  &  19      & 15.1 & \\
VXR70  & 10.85 & (0.64) & 0.75  & $-$0.9  &  17      & 13.4 & \\
VXR71A & 15.32 & ---    & 2.41  & +3.7    &23.5      & 20.9 & SB1?\\
VXR72  & 11.46 & (0.73) & 0.84  & $-$0.6  &  15      & 14.1 & \\
VXR76A & 12.76 & (1.05) & 1.24  & +0.3    &  8.      & 14.4 & \\
VXR77A & 9.91  & (0.50) & 0.60  & $-$1.4  &  95:     & 9: & \\
VXR78  & 10.44 & ---    & 0.73  & $-$0.9  &  50:   & 22.8 & SB1?\\
VXR80A & 11.98 & ---    & 1.04  & $\sim$0?& $\sim$150 & ---&  \\
 & & & & &  & & \\ \hline 
 & & & & &  & & \\ 
\multispan{8}{(1) Names and photometry 
were taken from Patten \& Simon (\cite{ps96}), who had retrieved 
B--V colors for VXR~7, VXR~30, \hfill}\\
\multispan{8}{and VXR~44 from the literature; the origin of B--V
colors for the other stars is not clear and they have not been used \hfill}\\
\multispan{8}{in the present
analysis (see text)\hfill}\\
\end{tabular}
\end{table*}
Lithium studies for IC~2602 and IC~2391 have previously been carried out by
R97 and Stauffer et al. (\cite{sta89},
hereafter S89), respectively.
Due to the relatively small size of their sample,
S89 could not conclude much about the presence of a spread
in IC~2391; nevertheless, their data allowed them to show
that the coolest stars in the cluster are significantly Li depleted
and that several Pleiades stars exist that have a higher Li 
content than IC~2391 members. 
R97 found that G and early--K
type stars in IC~2602 generally lie on the upper bound of the Pleiades 
Li vs. \teff~distribution, with only a
few early K--type stars showing signs of lithium depletion.
Their data evidenced some star-to-star scatter among late--G and
early--K dwarfs
in IC~2602, suggesting that the spread develops during the 
PMS phase.
More surprisingly, and similarly to the results of S89 for IC~2391,
R97 found that the latest--type stars
in IC~2602 were severely depleted in lithium and that their average lithium
abundance was comparable to the lowest abundances measured among the Pleiades.

The new lithium observations presented in this paper significantly enlarge
the two previous surveys and extend them to later spectral--types. 
The new data will allow us to verify with higher
statistical significance the results of S89 and R97 and to investigate
the Li vs. color distribution for very low mass stars
in the two clusters. The same spectra will also be used to ascertain 
membership and to derive rotational velocities for those IC~2602 
candidate members not included in the samples of S97
and R97. Besides lithium abundances, we also estimated
iron abundances for a subsample of stars in the two clusters. 
\section{Observations}
The observations were carried out at the European Southern Observatory
(ESO) and at the Cerro Tololo Inter--American 
Observatory (CTIO). The stars observed at CTIO are the ones 
already studied by S97 for membership, activity, and rotation.
The 4m telescope was
used, in conjunction with
the red, long--camera echelle spectrograph and a 31.6 line~mm$^{-1}$
grating. A 120$\mu$m slit width (0.8 arcsec on the sky) and a 
Tektronix 2048 $\times 2048$ CCD provided a 2 pixel resolution of $\sim 0.15$
\AA~at H$\alpha$ (R$\sim 43,800$) and a spectral coverage from $\sim$ 5800
to 8200~\AA.
Data reduction was performed using IRAF (see S97 for details).
Since radial and rotational velocities were already
derived by S97, we used their spectra only to carry out the lithium 
and metallicity analysis. We obviously excluded from the 
sample the stars which were found to be cluster non-members by S97.

Additional spectra for 15 X-ray selected IC~2602 candidate members
were obtained at ESO in April 1995, using CASPEC at the 3.6m telescope
in a similar configuration to the one used by R97.
Briefly, the standard
grating (31.6 lines/mm), together with the red cross-disperser,
the short camera, and ESO CCD \#32 (512 $\times$ 512 27 $\mu$ pixels)
were used. The slit width was 300 $\mu$ (2.25 arcsec on the sky),
resulting in a resolving power $R\sim 18,000$. 
The observed stars cover a large range of magnitudes and
the spectra were taken under various weather conditions; the achieved S/N
per resolution element thus ranges between $\sim 30$ and $\sim 100$. 
The data were reduced using the context {\sc echelle} within MIDAS;
the usual steps were followed; namely, background (including scattered light)
subtraction, flat-fielding,
order extraction, sky subtraction, and wavelength calibration.
We mention that the level of scattered light in CASPEC is very low ($<$ 3\%)
and it is well taken into account with a proper background subtraction.
Examples of the spectra
at lithium and H$\alpha$ are shown in Fig.~1. 

The sample stars are listed in Table~1;
three of the stars observed in the ESO '95 run (R~24A, R~31,
and R~38) were already observed at CTIO by S97 and, in turn,
three of the stars in the S97 sample (R~15, R~66, and R~70) had been observed
by R97. Radial and rotational velocities as well as H$_\alpha$ EWs are
also listed in the table. For stars included in S97 sample these quantities
were retrieved from that paper.
\section{Analysis}
\subsection{Rotational and radial velocities}
Radial and rotational velocities for the new sample of IC~2602
stars observed at ESO in April 1995 were derived via cross-correlation
techniques in the same manner as for the other IC~2391 and 2602
data (S97). We primarily used two orders
($\lambda\lambda$\ 6000-6070~AA\ and $\lambda\lambda$\ 6405-6485~AA)
dominated by a large number of modestly strong absorption
features. For the two very rapid rotators where these two
orders were less useful, we instead relied primarily on the
order which included the Na~{\sc i} D doublet. The radial velocity
zero point was set by observations of a number of rv standard
stars (specifically HD~102870, 126053, 154417 and
161096).  HD~161096 was used as the template star against which
the spectra of the IC~2602 program stars were cross-correlated.
An order dominated by O2 lines in the Earth's atmosphere was
used to determine small corrections to the radial velocity
zero point for each target object due to spectrograph flexure
(or other time or hour-angle dependent changes).
\begin{table*}
\caption{Li abundances for sample stars}
\begin{tabular}{lcrcccc} \hline
 & & & & & &\\
\multispan{4}{IC~2602\hfill}\\
 & & & & & &\\
star & \teff & $\Delta$\teff & EW(Li) & $\log$ n(Li)$_{\rm LTE}$& $\log$ n(Li)$_{\rm NLTE}$ &$\Delta \log$ n(Li) \\
      & (K)   & (K) & (m\AA) &    &     &    \\
 & & & & & & \\
W79  & 5260 & 57      & 142 $\pm$ 5 & 2.3 &    2.4      & 0.14 \\
R1   & 5050 & 89      & 204 $\pm$ 7 & 2.4 &    2.4      & 0.18 \\
R8A  & 5930 & 86      & 178 $\pm$ 9 & 3.2 &    3.1      & 0.20 \\
R15  & 4820 & 76      & 255 $\pm$15 & 2.4 &    2.4      & 0.24\\
R24A & 3780 & 100$^*$ & 112 $\pm$ 8 & 0.3 &    0.3      & 0.2$^*$ \\
R26  & 3540 & 100$^*$ & $\leq$ 100  & --- &  ---         & --- \\
R27  & 3830 & 100$^*$ & $\leq$ 50   & --- &  ---         & --- \\
R30  & 3440 & 100$^*$ & $\leq$ 80   & --- &  ---         & --- \\
R31  & 3480 & 100$^*$ & $\leq$ 55   & --- &  ---         & --- \\
R32  & 3540 & 100$^*$ & $\leq$ 60   & --- &  ---         & --- \\
R38  & 3330 & 100$^*$ & $\leq$ 50   & --- &  ---         & --- \\
R39  & 3310 & 100$^*$ & $\leq$ 80   & --- &  ---         & --- \\
R44  & 3630 & 100$^*$ & $\leq$ 70   & --- &  ---         & --- \\
R50  & 3590 & 100$^*$ & $\leq$ 50   & --- &  ---         & --- \\
R52  & 4580 & 53      & 262 $\pm$11 & 2.1 &     2.2      & 0.18 \\
R53B & 3340 & 100$^*$ & ---         & --- &  ---       & ---\\
R54A & 4260 & 93      & 189 $\pm$10 & 1.2 &  ---       & 0.21 \\
R57  & 3360 & 100$^*$ & $\leq$ 100  & --- &  ---         & --- \\
R66  & 5560 & 174     & 173 $\pm$10 & 2.8 &     2.8      & 0.3 \\
R67  & 3540 & 100$^*$ &  72 $\pm$12 & $-$0.3 &  ---      & 0.24$^*$ \\
R70  & 5760 & 62      & 172 $\pm$12 & 3.0 &     2.9      & 0.2 \\
R73A & 3340 & 100$^*$ & $\leq$ 60   & --- &  ---         & --- \\
R77  & 3910 & 100$^*$ & $\leq$ 30   & --- &  ---         & ---\\
R82  & 3330 & 100$^*$ & $\leq$ 56   & --- &  ---         & ---\\
R93  & 3980 & 10      &  60 $\pm$10 & 0.0 &  ---        & 0.12 \\
R96  & 4160 & 10      & 320 $\pm$12 & 2.0 &  ---         & 0.14  \\
& & &  & & & \\ \hline
& & &  & & & \\
\multispan{7}{Asterisks indicate stars for which errors in \teff~could
not be directly estimated\hfill}\\
\end{tabular}
\end{table*}
\begin{table*}
\begin{tabular}{lrccccc} \\ \hline
 & & & & & & \\
\multispan{4}{IC~2391\hfill}\\
 & & & & & & \\
star & \teff & $\Delta$\teff & EW(Li) & $\log$ n(Li)$_{\rm LTE}$ & 
$\log$ n(Li)$_{\rm NLTE}$ &$\Delta \log$ n(Li) \\
      & (K) & (K)  & (m\AA) &  & &      \\
 & & & & & & \\
 VXR3A  & 5590 & 100$^*$ & 175$\pm$ 16 & 2.9 & 2.8 & 0.27$^*$\\
 VXR6A  & 3690 & 34      & $\leq$ 60   & --- &  --- &---\\
 VXR7   & 6470 & 100$^*$ & 116$\pm$  9 & 3.3 & 3.2 & 0.21$^*$\\
 VXR14  & 5780 & 106     & 162$\pm$ 11     & 3.0 & 2.9 & 0.23 \\
 VXR16A & 4970 & 73      & 302$\pm$ 12 & 3.0 & 2.8 & 0.21 \\
 VXR18A & 3990 & 25      &  55$\pm$ 15 & 0.0 &  --- & 0.18  \\
 VXR30  & 6470 & 100$^*$ & 109$\pm$ 12 & 3.3 & 3.2 & 0.24$^*$ \\
 VXR31  & 5630 & 100     & 182$\pm$ 10 & 3.0 & 2.9 & 0.22\\
 VXR44  & 6650 & 100$^*$ &  71$\pm$ 12 & 3.0 & 2.9 & 0.24$^*$ \\
 VXR49B & 3730 & 33      & $\leq$ 50   & --- & --- & --- \\
 VXR50A & 5050 & 76      & 239$\pm$ 13 & 2.6 & 2.6 & 0.22 \\
 VXR62A & 4841 & 67      & 250$\pm$ 15 & 2.4 & 2.4 & 0.23\\
 VXR65  & 3820 & 30      & $\leq$ 45   & --- & --- & --- \\
 VXR67A & 4750 & 63      & 255$\pm$ 16 & 2.3 & 2.3 & 0.27 \\
 VXR69A & 5080 & 77      & 251$\pm$ 17 & 2.7 & 2.7 & 0.25 \\
 VXR70  & 5557 & 97      & 182$\pm$ 10 & 2.9 & 2.8 & 0.22 \\
 VXR71A & 3360 & 15      & $\leq$ 70   & --- & --- & --- \\
 VXR72  & 5257 & 85      & 225$\pm$ 13 & 2.8 & 2.7 & 0.22\\
 VXR76A & 4343 & 43      & 204$\pm$ 11 & 1.4 & --- & 0.17 \\
 VXR77A & 6150 & 120     &  90$\pm$  8 & 2.9 & 2.8 & 0.22 \\
 VXR78  & 5630 & 100     & 178$\pm$ 10 & 2.9 & 2.8 & 0.22 \\
 VXR80A & 4720 & 61      & 232$\pm$ 34 & 2.3 & 2.2 & 0.37 \\
& & &  & & &\\ \hline
\end{tabular}
\end{table*}
The derived rotational and radial velocities are listed in
Cols.~6 and 7 in Table~1. Based on multiple spectra of
some of the radial velocity standards, we believe the one
sigma uncertainty for the IC 2602 radial velocities should
be $\pm$2~km/s for stars with v$\sin <$ 30 km/s, increasing
up to several km/s for the most rapid rotators (stars with
expected uncertainties $>$~2 km/s are marked with a colon
after the radial velocity in Table~1).  Based on our previous
experience with the cross-correlation technique, we expect
the rotational velocities to have uncertainties of order
10\% of the vsini.

Two or more radial velocity standards were observed on
each night of the run, with a total of 10 spectra for
the rv standards obtained during the run.  We can estimate the 
external accuracy (relative to these radial velocity standards) for
our derived radial velocities from the RMS scatter of the
velocities we derive - that RMS is 1.7 km/s.  Comparison
of the new observations versus that in S97 for the four
stars in common suggests that new radial velocities may be
low by 1-2 km/s, but that comparison is made less useful by
the large average rotational velocities for these four stars.
\subsection{Effective temperatures and Li abundances}
Effective temperatures were derived using \bmv~and \vmi~ color
indices 
and assuming reddening values 
E(B$-$V)=0.04 and E(V$-$I)=0.04 for IC~2602 (Whiteoak \cite{whi61};
Braes \cite{bra62})\footnote{
We used these reddening values for consistency with both
S97 and R97. Note, however, that if using E(B$-$V)$=0.04$,
a reddening E(V$-$I)$=0.05$
would be more appropriate 
(see Pinsonneault et al. \cite{pin98}). Using the latter,
we would have obtained $\sim$ 20~K warmer temperatures with negligible
differences in the inferred \nli~and [Fe/H] values.}
and E(B$-$V)=0.01 and E(V$-$I)=0.01 for IC~2391 (see discussion in
Patten \& Simon \cite{ps96}).
For stars with (B--V)$_0 \leq 1.4$ and both color indices available, we
used both color calibrations and then we adopted the average \teff~value;
on the other hand,  
for IC~2391 stars observed at CTIO one color only is available, either
B--V (for the three warmest stars) or V--I\footnote
{Patten \& Simon (\cite{ps96})
list in their Table~4 (see also Table~1 in the present paper) 
B--V colors for several
of our sample stars; however, we decided not to use those colors since
{\it a)} the origin of the data is not clear (Patten \& Simon carried
out $V,~R,$ and $I$ photometry only); {\it b)} B--V colors seem too blue
providing
systematically higher \teff~than V--I colors.}; for these stars
we were forced to assume \teff$=$\teff(B--V) or \teff$=$\teff(V--I). Finally,
for stars with (B--V)$_0>1.4$ in both clusters
we used only effective temperatures
derived from V--I colors,
since B--V is not a good temperature indicator for very cool stars.
Note that since the B--V and V--I calibrations are based on Bessel 
(\cite{bes79})
the use of both calibrations for part of the stars and only one for
the others should not introduce any systematic temperature difference.
As to the color--temperature calibrations,
we used the \teff~vs. \bmv~calibration of Soderblom et al. 
(\cite{sod93})\footnote{\teff(\bmv)=1808(B--V)$^2_0-6103$(B--V)$_0+8899$} and,
for stars with \vmo$\leq 1.6$, the \teff~vs. \vmi~calibration of R97\footnote
{\teff(\vmi)=$-755$(V--I)$_0^3+4246$(V--I)$^2_0-8598$(V--I)$_0+9900$};
as mentioned, both calibrations are based on Bessel (\cite{bes79}). 
For stars redder than
\vmo$=1.6$ we instead used the \teff~vs. \vmi~calibration of Bessel 
(\cite{bes91}),
which is about 200~K cooler than the one by 
Kirkpatrick et al. \cite{kir93}) 
and between $\sim$100 and 150~K cooler than the
one by Leggett et al. (\cite{leg96}). We have employed this calibration for
consistency with the calibration used for warmer stars; although this 
may introduce systematic errors in our results, it will not affect
the relative comparison between different samples analyzed in the same way.

Patten \& Simon (\cite{ps96}) 
estimate an error of $\sim$~2\% in their photometry
which reflects into an error $\delta$(\vmi)$=0.028$~mag;
we derived the error in the inferred effective temperatures accordingly.
On the contrary, no estimates of photometric errors are provided by
Prosser et al. (\cite{pro96}) for IC~2602 stars; they just mention that
for stars brighter than V$=16$ ``the accuracy of the photometry
is generally limited by variability" (due to rotational modulation
of starspots). Since for these stars we found
no systematic difference between \teff(B--V) and \teff(V--I),
we estimated 
random errors in \teff~as $\delta$\teff=
$0.5\times\sqrt{\Delta \rm T_{\rm eff}^2 }$, with $\Delta$\teff=
\teff(B--V)$-$ \teff(V--I).
Finally, for stars for which \teff~was derived from one color only
and photometric
errors are not available,
we assumed a constant error of 100~K,
slightly larger than the average 
error for the stars for which \teff~was derived by averaging the values 
given by both indices. 
The adopted temperatures and
errors are listed in Cols.~2 and 3 of Table~2.

At the resolution of our observations (both the ESO and CTIO ones)
the Li feature is blended with 
lines of other atomic and molecular species, the main contributor
being the Fe~{\sc i}~$\lambda$~6707.44~\AA~line.
We used the relationship given by Soderblom et al. (\cite{sod93}) to estimate
the contribution of the Fe line as a function of \bmv; we then derived 
a EW(6707.44) vs. \teff~relationship, which we used to estimate
the contribution of the Fe line for stars with no available \bmv. 
For stars cooler than about 3900~K the formula given by Soderblom et al. 
(\cite{sod93}) is not appropriate, 
as, on the one hand the Fe~{\sc i} line
is not as strong as in warmer stars and, on the other hand, the contribution
of molecular lines becomes more significant.
Among very cool stars (\teff$\leq 3900$~K) we have only two Li
detections and a somewhat inhomogeneous set of upper limits; since for
these stars we 
will investigate the behaviour of lithium
directly on the EW vs. V--I plane, \nli~are computed only for the two
stars in IC~2602 with a Li detection. For them we assumed
an equivalent width of the 6707.44~\AA~
line of 20~m\AA, as measured in the spectrum of two field M dwarfs with similar
colors (Gl 193 and Gl 273). We caution that the Li
abundances we provide for these two stars of IC~2602 should be regarded 
as indicative only.

LTE lithium abundances were then derived using corrected equivalent widths and
the curves of growth (COGs) of Soderblom et al. (\cite{sod93}). 
These COGs do not 
extend above 6500 K and below 4500 K; for stars warmer and cooler than
these temperatures we estimated \nli~by extrapolating the COGs, as was done by 
Soderblom et al. (\cite{sod99}). 
LTE abundances of stars warmer than 4500~K were
corrected for NLTE effects using the code of Carlsson et al. (\cite{car94}).
NLTE corrections are not 
provided in that paper for stars cooler than 4500~K and thus no correction was 
applied to those stars. NLTE corrections for a larger
temperature range (down to 2500~K) have been computed by Pavlenko et al. 
(\cite{pav95}), but unfortunately they are not provided in tabular form. 
In any case, as discussed in detail in that paper,
NLTE corrections become smaller
as one moves to cooler temperatures.
\begin{table*}
\caption{Li abundances for the S89 sample}
\begin{tabular}{lcrcccc} \hline
 & & & & & &\\
star & \teff & $\Delta$\teff & EW(Li) & $\log$ n(Li)$_{\rm LTE}$ &
$\log$ n(Li)$_{\rm NLTE}$ &$\Delta \log$ n(Li) \\
      & (K) & (K)  & (m\AA) & &  &     \\
 & & & & & &\\
SHJM~1 & 5590 & 95  & 160  $\pm$ 10 & 2.8 & 2.7 & 0.22\\
SHJM~2 & 5970 & 147 & 140  $\pm$ 10 & 3.1 & 2.9 & 0.27\\
SHJM~3 & 4480 & 166 & 130  $\pm$  7 & 1.2 & --- & 0.77\\
SHJM~6 & 5050 & 75  & 190  $\pm$ 11 & 2.3 & 2.4 & 0.20\\
SHJM~7 & 4380 & 36  & $\leq$ 50  & $\leq$ 0.39& --- & ---  \\
SHJM~8 & 4020 & 98  & 90   $\pm$ 30 & 0.4 & --- & 0.38\\
SHJM~9 & 4030 & 87  & 100  $\pm$ 30 & 0.4 & --- & 0.37\\
SHJM~4 & 3558 & 20  & $\leq$ 50  & $\leq$ 0.& --- & ---  \\
SHJM~5 & 3893 & 28  & $\leq$ 200 & $\leq$ 1.1& --- & ---  \\
SHJM~10& 3580 & 20  & $\leq$ 100 & $\leq$ 0.46& --- & ---  \\
& &  &&  & &\\ \hline
\end{tabular}
\end{table*}
\begin{table*}
\caption{Recomputed Li abundances for R97}
\begin{flushleft}
\begin{tabular}{rcccc}  \hline
  &   &     & & \\
star & \teff (R97)& \nli (R97) & \teff (new) & \nli (new) \\ 
     & (K) & & (K) & \\ 
  &   &   & & \\
R~ 3~~&5080& 2.74     & 5140 &2.79\\
R~ 7~~&6750& $>$ 2.97 & 6860 & 3.14\\
R~14~~&5080& 2.58     & 5150 & 2.65 \\
R~15~~&4900& 2.55     & 4820 & 2.44 \\
R~21~~&6430& 3.18     & 6320 & 3.10  \\
R~29~~&4440& 1.76     & 4440 & 1.77 \\
R~35~~&5770& 3.12     & 5820 & 3.12 \\
R~43~~&4840& 2.99     & 4740 & 2.89 \\
R~45~~&5810& 2.87     & 5800 & 2.87 \\
R~56~~&3910& $\leq$ 0.65& 3990 & 0.64 \\
R~58~~&5850& 3.26 (3.12) & 5780 &3.25 (3.06) \\
R~59~~&5240& 3.09     & 5060 & 2.91 \\
R~66~~&5730& 2.97     & 5560 & 2.83 \\
R~68~~&5110& 2.77     & 4880 & 2.57 \\
R~70~~&5700& 2.85     & 5760 & 2.91 \\
R~72~~&5890& 3.19     & 5770 & 3.09 \\
R~79~~&6750& $>$2.94  & 6730 & 3.02 \\
R~83~~&5970& 3.22     & 5770 & 3.07 \\
R~85~~&6390& 3.06     & 6390 & 3.06 \\
R~89~~&4180& 1.60     & 4180 & 1.61 \\
R~92~~&5690& 2.98     & 5630 & 2.93 \\
R~94~~&3950& $\leq$ 0.87 & 3903 & 0.77  \\
R~95A~&5080& 2.99     & 5020 & 2.93 \\
&    &   &   &  \\ \hline
\end{tabular}
\end{flushleft}
\end{table*}
Measured EWs (not corrected for the contribution of the $\lambda 6707.44$~\AA~
feature), with errors are listed in Col.~4 in Table~2.
Derived Li abundances (both in LTE and NLTE)
and errors are listed in Cols.~ 5, 6, and 7. 
In the following (and in the figures)
NLTE abundances will be used when available. Note that the differences between
LTE and NLTE abundances are within the errors in inferred \nli~and thus
using NLTE abundances for a subsample of the stars and LTE abundances
for the other subsample should not introduce any bias.
Errors in \nli~include the contribution of errors in \teff~and of errors
in equivalent widths to which we added 5~m\AA~to take into account
possible errors in the estimate of the atomic$+$ molecular blend contribution.
\subsection{Reanalysis of previous Li observations}
Effective temperatures and lithium abundances (together with their errors)
were recomputed for the R97 IC~2602 sample as 
well for the S89 IC~2391 sample in the same fashion as for our new 
sample, using the published EWs and photometry. More specifically, photometry
for IC~2602 stars was taken from Prosser et al. (\cite{pro96}), while photometry
for IC~2391 stars was retrieved from S89 (B--V colors) and Patten \& Simon
(\cite{ps96} ~~-V--I colors). Li EWs were taken from R97 and S89 for IC~2602
and IC~2391, respectively. R97 had derived \nli~adopting 
the same COGs, but they assumed 
\teff$=$\teff(\bmv) and did not extrapolate the COGs for stars warmer
than 6500 and cooler 4500~K.
S89 provided only EWs and did not compute Li abundances.
Li abundances for the S89 stars are listed in Table~3, while 
``new'' and ``old'' Li abundances and \teff~ for the R97 sample are listed
in Table~4. Note that SHJM7 was not confirmed as a cluster
member by Patten \& Simon (\cite{ps96}); therefore, although we list it
in Table~3, it is not considered in the following discussion 
and not plotted in the figures.
Table~4 shows that 
the agreement between the ``old'' and ``new'' Li abundances for
the sample of R97 is quite good. In the following these 
reanalysed Li abundances from S89 and R97 will be combined with our new 
data to investigate the behaviour of lithium in the two clusters.
\subsection{Fe abundances} 
Iron abundances $\log \epsilon$(Fe)
were derived using a selected sample of relatively high S/N
spectra in IC~2602 (nine stars, mainly coming from the R97 sample) 
and IC~2391 (four stars). 
We used the EWs of 4--8 Fe~{\sc i} lines (not all the lines were measurable
in all the spectra),
whose wavelengths are listed in 
Table~5; the EWs of the lines can be made available by one of us (SR)
upon request. A first very qualitative estimate of the metallicities
of the two clusters was carried out by plotting the EWs of these iron lines
as a function of effective temperature. The corresponding diagrams are
shown in Fig.~2, where the two Pleiades stars analyzed by
King et al. (\cite{kin00a}) are also included. A large spread in EWs is present
for most of the lines for both the IC clusters and the two Pleiades
points. 
Focusing on the lines affected by a relatively small amount of scatter,
the figure suggests that the two IC~clusters have similar metallicities
and that their Fe abundances do not differ much from that of the Sun
and of the Pleiades.

A more quantitative metallicity estimate was carried out starting from the EWs
of the iron lines shown in Fig.~2 and
using the same line-to-line abundance code 
described by Carretta \& Gratton (\cite{cg97}). Kurucz (\cite{kur95}) 
model atmospheres,
including overshooting, were used.
gf-values for the eight Fe~{\sc i} lines were determined from an inverse
abundance analysis of
the same iron lines in
the solar spectrum, using the same code and model atmospheres and assuming
a solar iron abundance $(\log \epsilon$(Fe)$_{\odot}$)=7.52).
Solar EWs, 
measured on the spectrum of the solar flux, were retrieved
from King et al. (\cite{kin00a}); 
the same solar parameters as King et al. (\teff$_{\odot}=5770$~K,
$\log g_{\odot}=4.44$, $\xi_{\odot}=1.1$~km/sec)
were assumed; note that microturbulence for the Sun is on the same
scale as for our sample stars (see below).
Finally, Van der Waals broadening was treated using the Uns\"old
(\cite{uns55}) approximation and the same enhancement factors as Gratton \&
Sneden (\cite{gs91}).
We mention that the spectrum of the solar flux has 
a higher resolution than our target spectra; furthermore,
as described in Sect.~2, within our sample itself, spectra taken in different
runs have different resolutions. In particular the spectra of the four
IC~2391 stars have been obtained at CTIO and have higher resolution than
those obtained at ESO. However, the use of spectra with different
resolution should not significantly affect our results 
since we have chosen spectral lines which are free from
significant blends\footnote{$\sim 0.2$~\AA~blueward of the Fe~{\sc i}
6703.57~\AA~feature
a very weak CN line is present; this line is blended with the Fe~{\sc i}
line at the resolution of our spectra. However the EW of the 6703.57~\AA~line
listed by King et al. (\cite{kin00a}) also includes the contribution of the CN feature
($\sim 1$~m\AA~in the Sun) 
and thus the gf-value inferred by an inverse abundance analysis
takes into account
both lines.};
in order to confirm this point, we have compared EWs
of the stars in common between the CTIO and ESO samples (see Sect.~3.5)
and found that differences are within measurement errors and, more important,
do not show any systematic trend.

For the analysis of our sample stars we adopted the same \teff~used for
inferring Li abundances,
a surface gravity $\log \rm g=4.5$, and a microturbulence velocity given by 
$\xi=3.2 \times 10^{-4}~(\rm T_{\rm eff}-6390)-1.3(\log\rm g-4.16)+1.7$
(see Boesgaard \& Friel \cite{bf90}).
In Table~5 we list the inferred
Fe abundances for each line and the 
mean abundance that was derived
using all the available lines for each star and giving the same weight
to each line. Estimates of the internal 
errors on the final abundance, inferred as explained in the Appendix,
are also listed in the table.
In order to put the Pleiades on the
same metallicity scale as our stars, we derived $\log \epsilon$(Fe) values for
the two Pleiades stars hii~97 and
hii~676 analyzed by King et al. (\cite{kin00a}). Stellar parameters
were computed as for our sample stars and the same metallicity analysis
was carried out. The results are also listed in Table~5. 
It is important to stress that,
as discussed in the Appendix, external uncertainties are not easy to estimate,
but can be as large as $\sim 0.1$ dex; although as in other studies of
this kind we cannot provide an absolute iron abundance,
the following discussion will be based on relative metallicities
(both between the two IC~clusters and between them and the Pleiades)
and thus should not be affected by external errors.
\subsection{Comparison of stars in common between different samples}
The stars in common between different samples, together with Li and 
H$\alpha$ EWs measured in the ESO and CTIO spectra, are listed in Table~6.
Inspection of the table shows a very good agreement as far as lithium is
concerned. In the following, when considering the merged sample, 
we will use \nli~derived from our new ESO $+$ CTIO sample and, because
of their higher S/N, we 
will give preference to the ESO data when both ESO and CTIO data 
are available (but using the other data would make no difference).
As to H$\alpha$, a large discrepancy between the ESO and CTIO values is seen
only for R~38. 

The CTIO spectrum of R~38 obtained in January 1995 showed a
peculiar H$\alpha$ profile with a relatively normal core plus
unusually broad and strong ``wings" (with about half of
the equivalent width being attributable to the broad
component). S97
discussed this profile in detail, and speculated that the star
had been observed shortly after a flare. The ESO 
spectrum of this star obtained in April 1995 was intended,
in part, to test that idea.  Because the new spectrum both
lacks the broad wings and has an equivalent width about half
of that observed four months earlier (see Fig.~1), we believe the flare
explanation is supported by the new data.
\section{Results}
\subsection{Cluster membership}
We have attempted to classify the stars we observed at ESO in 1995
as either members or non-members based on all of the information
available to us.   By selection, all of the stars are possible
members based on their photometry.  We have used the following
criteria as membership indicators, in approximately this order
of precedence: \begin{itemize}

\item A radial velocity within two $\sigma$ of the mean radial velocity
for the cluster, which we take to be v$_{\rm rad} \sim 16$~km/s for IC~2602.
This criterion may be violated for cluster members that are 
spectroscopic binaries;
\item spectral features appropriate for the observed colors of the
star assuming the nominal reddening to IC~2602;
\item chromospheric activity (as deduced from the H$\alpha$ equivalent
width and profile) consistent with the star's color and with the
age of IC~2602;
\item an inferred lithium abundance roughly compatible with the star's
mass and the age of IC~2602.
\end{itemize}

A membership flag based on the combination of these criteria is
indicated in the last column of Table~1. Our reasoning for the
stars not designated as members is as follows (except for R~28A
which is treated separately in the next section):
{\it R~8A --} the radial velocity is much higher than expected for a cluster
member.  However, all of the other spectral characteristics are
compatible with the star being an IC~2602 member, and we therefore
consider this star to be a possible SB1 cluster member;
{\it R~37A --} the very high rotational velocity for this star in combination
with its M dwarf colors suggests that this is a very young,
low mass star, and hence a likely IC~2602 member.  However, a
consequence of the rapid rotation is that we
are unable to measure a usefully accurate radial velocity 
or to detect lithium in absorption. H$\alpha$ is only weakly
in emission, whereas we would expect a strong H$\alpha$ emission
line given the star's rapid rotation and late--type colors.
The data for this star are thus contradictory, and we consider
its membership status unknown;
{\it R~42C --} we consider this star to be a non-member because its radial
velocity is far removed from the cluster mean and because it
does not have a detected lithium absorption feature despite
being in a color range where we would expect to easily detect
lithium even in stars several times older than IC~2602;
{\it R~46 --} this star also has a radial velocity far from the cluster mean
and a lithium equivalent width much less than we would expect
for its color;
{\it R~73  --} the large rotational velocity (or velocities) and late-type
colors strongly suggest this is a young star, and hence a
likely member of the cluster. The radial velocity is in
agreement with that assessment. We note that the H$\alpha$
profile and the profiles of a few of the stronger
absorption features look like there is a narrow core and
broader wings. This could indicate that the star is an
SB2, with one relatively slow rotator and one rapid rotator
(vsini $\sim$~30 and 90 km/s, respectively).   

All together (i.e., including R97 and S97 samples) 55 new IC~2602 candidates
from the X-ray survey were observed: 41 of them (74 \%) were confirmed as
members, 9 (18 \%) were rejected as members, and for the remaining five
additional spectra should be taken (but three of these five are very likely
SB cluster members as indicated by lithium). We conclude therefore that the
rate of success of the X--ray survey in identifying new cluster members is 
rather high, while the contamination by late--type field stars is
consistent with the prediction of Randich et al. (\cite{R95}). We
also note that, as expected, most of the non--members are found among
very late--type stars.
\subsubsection{A peculiar non--member: R~28A}
We are uncertain how to interpret the spectrum of one of
the candidate IC~2602 members observed at ESO in 1995.
The star in question is R~28A.  This star has V = 15.68,
\bmv = 2.07, \vmi = 2.53. There are at least two indications
in the spectrum that it is a non-member: (a) the derived
radial velocity is +97~km/s 
and (b) whereas all other cluster
members with \vmi~$>$ 2.0 show TiO absorption features
in our spectra, R~28A does not, and instead has a spectrum
more compatible with that of an early K star.  These two
characteristics might suggest that R~28A is instead a
background, K giant.  However, R~28A also has a very strong
Li~6708~\AA\ absorption feature, with an equivalent width
of ~530 m\AA~ - perhaps suggesting that the object is a
distant, pre-main sequence K star.  The H$\alpha$ profile of
the star is also peculiar, since it includes both emission
and absorption components (see Fig.~1).  The emission
component is suspiciously ``at rest" with an observed
wavelength of 6562.9~\AA, whereas the absorption component
has exactly the wavelength expected for the radial
velocity derived for R~28A using the atomic absorption lines
in two of the orders blueward from H$\alpha$.   While this star
seems to have many interesting characteristics, enough of
those characteristics indicate that it is very unlikely to
be a member of IC~2602 and we can safely exclude it from
further discussion here.
\subsection{Metallicity}
In Fig.~3 we show the average $\epsilon$(Fe) as a function
of \teff~ for IC~2602 and IC~2391 
stars listed in Table~5. The two Pleiades stars reanalyzed by us are also shown.
The horizontal lines indicate the weighted mean for the three clusters,
\efe$=7.47 \pm 0.05$ for IC~2602, \efe$=7.49 \pm 0.07$ 
for IC~2391, and \efe$=7.49 \pm 0.1$ for the Pleiades.
Weighted means were computed assuming for each star a conservative final
internal error $\sigma=\sigma_1+\sigma_2$.
\tiny
\begin{table*}
\caption{Metallicity}
\begin{tabular}{rllccccccccc}  \hline
  &   &   & & &    & && & & &\\
name & \teff & $\xi$ & \multicolumn{5}{c}{$\log \epsilon$(Fe)} &\\
 & (K) & (km/s) & 6703.57 & 6710.32 &
6725.36 & 6726.67 &
 6733.16 & 6750.16 & 6810.27 &
6858.15 & mean $\pm \sigma_1 \pm \sigma_2$\\ 
 & &  & &  &  & & &  & & & \\
\hline
R~1  & 5050 & 0.8 & 7.47 & ---  & ---  & 7.41 & 7.39 & 7.43 & 7.55 & -- &
7.45$\pm 0.06\pm 0.09 $ \\
R~14 & 5150 & 0.9 & 7.50 & ---  & ---  & 7.41 & 7.36 & 7.53 & 7.53 & 7.40 &
7.45$\pm 0.07\pm 0.08 $ \\
R~15 & 4810 & 0.7 & 7.57 & ---  & 7.48 & 7.30 & 7.47 & 7.26 & 7.59 & 7.60 &
7.44$\pm 0.14\pm 0.08$ \\
R~21 & 6320 & 1.2 & 7.51 & ---  & ---  & 7.32 & ---  & 7.50 & ---  & ---  &
7.44 $\pm 0.11\pm 0.1 $ \\
R~29 & 4440 & 0.6 & 7.53 & 7.50 & ---  & 7.52 & 7.50 & 7.39 & 7.65 & ---  &
7.51$\pm 0.06\pm 0.06 $ \\
R~66 & 5560 & 1.0 & 7.48 & ---  & ---  & 7.53 & 7.48 & 7.50 & 7.51 & 7.30 &
7.47$\pm 0.08\pm 0.13 $ \\
R~92 & 5630 & 1.0 & 7.49 & 7.51 & ---  & 7.49 & ---  & 7.64 & 7.47 & 7.50 &
7.52$\pm 0.06 \pm 0.07 $ \\
R~95 & 5020 & 0.8 & 7.51 & 7.50 & ---  & 7.44 & 7.39 & 7.45 & 7.47 & 7.28 &
7.43$\pm 0.08 \pm 0.07 $ \\
W~79 & 5260 & 0.9 & 7.52 & 7.42 & 7.52 & 7.47 & 7.39 & 7.49 & 7.48 & 7.38 &
7.46$\pm 0.05 \pm 0.07 $ \\
VXR~16A & 4970 & 0.8 & 7.51 & --- & ---  & 7.55 & ---  & 7.51 & 7.55 & ---  &
7.53$\pm 0.02 \pm 0.09 $ \\
VXR~67  & 4750 & 0.7 & 7.30 & --- & 7.51 & 7.55 & 7.50 & 7.36 & 7.51 & 7.40 &
7.45$\pm 0.09 \pm 0.09 $ \\
VXR~72  & 5260 & 0.9 & 7.39 & --- & ---  & 7.50 & ---  & 7.35 & 7.46 & ---  &
7.42$\pm 0.07 \pm 0.09 $ \\
VXR~76  & 4340 & 0.6 & 7.57 & 7.41 & ---  & 7.51 & ---  & 7.43 & 7.63 & 7.57 &
7.52$\pm 0.09 \pm 0.09 $ \\
hii~676  & 4400 & 0.6 & 7.57 & 7.44 & 7.57 & 7.46 & 7.44 & 7.44 & ---  & 7.43 &
7.48$\pm 0.06 \pm 0.06 $ \\
hii7  & 4445 & 0.6 & 7.56 & 7.37 & 7.59 & 7.58 & 7.54 & 7.37 & ---  & 7.49 &
7.50$\pm 0.09 \pm 0.07 $ \\
 & & & & &   &   & &  & & &\\
\hline
\end{tabular}
\end{table*}
\normalsize
\begin{table*}
\caption{Comparison of stars observed in different runs}
\begin{tabular}{ccccr} \hline
 & & & &\\
star & EW(Li) (ESO '94) & H$\alpha$ (ESO '94) & EW(Li) (CTIO) & H$\alpha$ 
(CTIO)\\
   & (m\AA) & (\AA) & (m\AA) & (\AA) \\
 & & & &\\
R~15 & 259 & $+$0.20 & 255 & $+$0.2 \\
R~66 & 186 & $-0.95$ & 173 & $-1.0$ \\
R~70 & 168 & $-1.05$ & 172 & $-1.2$\\ 
 & & & &\\
star & EW(Li) (ESO '95) & H$\alpha$ (ESO '95) & EW(Li) (CTIO) & H$\alpha$ 
(CTIO) \\
   & (m\AA) & (\AA) & (m\AA) & (\AA) \\
 & & & &\\
R~24A & 112 & $+$3.0 & 107 & $+1.9$\\
R~31  & $\leq 55$ & $+4.0$ & $\leq 70$ & $+3.9$\\
R~38  & $\leq 50$ & $+4.9$ & $\leq 100$ & $+11.3$\\
 & & & &\\ \hline
\end{tabular}
\end{table*}
The figure exhibits a spread in Fe abundances, but
the star-to-star scatter is well within the formal error bars.
A larger number of stars and, in particular, higher S/N spectra that
would allow measuring the EWs of a larger set of Fe lines,
are certainly desirable for a more accurate abundance analysis;
our data do not allow us to ascertain whether there
is some {\it real} star-to-star scatter in \efe, neither
can we infer an absolute iron abundances for the IC~clusters;
nevertheless, we think that the claim
can be made that the mean abundances of the two clusters are similar
(with IC~2391 possibly slightly more metal rich)
and that both are close to the solar abundance
 (contrary to preliminary
report by Meola et al. \cite{meo00} of somewhat subsolar abundances based on a
less accurate abundance analysis). 
Assuming a solar abundance $\epsilon$(Fe)$_{\odot}=7.52$, the mean abundances
for the two IC~clusters translate into mean [Fe/H] values $-0.05 \pm 0.05$
and $-0.03 \pm 0.07$, respectively. Their metallicity, therefore, is consistent
with the solar metallicity of most of the young clusters
for which spectroscopic abundances have been obtained so far.

The canonical value for the Pleiades derived by Boesgaard \& Friel 
(\cite{bf90}) is [Fe/H]$=-0.034\pm 0.024$
or \efe$=7.486$; however, as discussed by King et al. (\cite{kin00a}) 
who derived
a metallicity [Fe/H]$=+0.06\pm 0.05$, [Fe/H] spectroscopic determinations
for this cluster range between [Fe/H]$=-0.03$ and [Fe/H]$=+0.13$. Our own
determination is indeed comparable with that of Boesgaard \& Friel 
(\cite{bf90}),
but is almost 0.1~dex below that of King et al. (\cite{kin00a}). Most important,
having put the Pleiades and the IC clusters on the same abundance scale,
we are able to say that the metallicities of the two IC~clusters are
not significantly higher (or lower) than the Pleiades.
\subsection{Lithium abundance}
\subsubsection{G and K--type stars}
The usual \nli~vs. \teff~plot for stars warmer than \teff$=3800$~K
in the two clusters is shown in Fig.~4. Similarly to Fig.~6 of R97,
we show two lines indicating the range of values for the initial Li abundance
for Pop.~I stars as indicated by NLTE analysis of
pre-main sequence stars (e.g., Mart\'\i n et al. \cite{mar94})
and meteorites.
Only confirmed clusters members, as from Table~1, are considered in
the figure.
A polynomial regression (grade 4) of the observed \nli~vs. \teff~
distributions down to \teff$=3800$~K was carried out (rms values
equal to 0.3 and 0.15~dex were obtained for IC~2602 and IC~2391,
respectively); the regression
curves are also shown in the figure.
The \nli vs. \teff~distributions of solar--type stars in the two IC~clusters
are very similar. In particular,
stars warmer than $\sim$ 5800~K in both clusters do not show any
significant Li depletion and their abundance is consistent
with the initial value for Pop. I stars. As mentioned in the Introduction,
this point has already been discussed
in several previous studies which agreed in saying that
it implies that, whereas theoretical models predict that
solar--type stars should undergo PMS Li depletion,
observations show that no PMS Li depletion occurs in stars more massive
than $\sim$ 1~M$_{\odot}$. 
The apparent scatter among the hot stars is
within the errors and thus no real scatter can be inferred from the present
data. Below 5800~K, Li depletion starts being seen and
the average Li abundance, as shown by the regression curves,
is about the same for the two clusters down to about 5000~K.
For stars cooler than this, the average abundance of IC~2602 stars 
is higher than that of
IC~2391, although IC~2602 stars as depleted as IC~2391 stars of similar
temperature are present.
In addition, IC~2602 members exhibit a certain amount of star-to-star scatter;
more specifically, the dispersion is modest (but present)
for early--K stars around 5000~K and it grows larger for cooler stars.
The scatter among IC~2391 stars seems instead smaller.
In order to perform a more quantitative analysis of the presence(absence)
of a dispersion in lithium among late--G/early--K cluster members, we
carried out a polynomial regression (grade 2) of the EW(Li) vs V--I color
distributions for stars with 0.8 $\leq$\vmo$\leq 1.05$.
The amount of dispersion can be estimated using a $\chi^2$-like quantity:\\
$\chi_{\rm disp.}=\frac{1}{\rm N}\times\sum{[(EW(Li)-reg)/(\Delta~EW(Li)]^2}$,\\
where $\rm reg$ is the value of the regression and N the number of stars. 
We found $\chi_{\rm disp} \sim 11$
for IC~2602 and $\chi_{\rm disp} \sim 6$ for IC~2391. 
In order to compare
the dispersion observed in the Pleiades with the two IC~clusters,
we retrieved Li EWs and colors for the Pleiades stars from the literature.
More specifically, lithium data 
were taken from Soderblom et al. (\cite{sod93}),
Garc\'\i a L\'opez et al. (\cite{gar94}), and Jones et al. (\cite{jon96}).
For the stars in common between Jones et al. (\cite{jon96}) and Garc\'\i a
L\'opez et al. (\cite{gar94}) we used Jones et al. data since they are
of higher quality; for the stars in common between Garc\'\i a
L\'opez et al. (\cite{gar94}) and Soderblom et al. (\cite{sod93}), we used
Soderblom et al.'s measurements for stars with quality code ``a"
in Soderblom et al. 
Table~1, and Garc\'\i a L\'opez et al. data in all the other
cases. (\vmi)$_{\rm K}$ colors were instead
retrieved from the Open Cluster Database\footnote{Open
Cluster Database, as provided by C.F. Prosser (deceased) and J.R. Stauffer,
and which currently may be accessed at 
ftp://cfa-ftp.harvard.edu/pub/stauffer/clusters/,
or by anonymous ftp to cfa-ftp.harvard.edu, cd /pub/stauffer/clusters.}
and converted into (\vmi)$_{\rm C}$ colors using the relationship of
Bessel \& Weis (\cite{bw87}). Errors in the EWs of the Li line are provided
for most of the stars in the sample of Garc\'\i a L\'opez et al.
(\cite{gar94}) and they are in the range $\sim 5-25$~m\AA. 
Jones et al. (\cite{jon96}) quote errors of 5 and 20 m\AA~for
slow and rapid rotators, respectively. Finally, Soderblom et al. (\cite{sod93})
quote an average error in their Li EWs of the order of 15~m\AA. In other words,
the quality of the Pleiades data is comparable with that of the data
presented here.
Using the expression above and the errors quoted in the three Pleiades papers,
we obtained for this cluster $\chi_{\rm disp}=16.5$, suggesting that
the dispersion is indeed larger than for the two IC~clusters.
Note however that whereas the two IC~samples in this color 
range are similar in size
(seven and eight stars, respectively), the Pleiades sample is much larger;
therefore, we cannot exclude that small number statistics
is the reason for the smaller scatter measured for the
IC~clusters.
\subsubsection{M dwarfs}
Plotted in Fig.~5 are the measured 
equivalent widths (EWs) of the 6707.8 \AA~lithium
line as a function of de-reddened \vmi~ color for late--K and
M cluster members with (V--I)$_0 > 1.3$ 
(which corresponds to \teff $<$ 4240~K in our calibration);
the coolest stars shown in Fig.~4 are also included in Fig.~5
and fall in the color interval $1.2 < (\rm V-\rm I)_0 < 1.7$.
As already noted in the previous section,
IC~2391 late--K stars have, on average, less Li than stars of the same
color in IC~2602 and do not display any major dispersion in Li.
This is still true as we move to very late spectral--types; in particular,
the lithium feature is not detected in IC~2391 members later
than (V--I)$_0 \sim$ 1.5, whilst in IC~2602 a star as late as
R~67 (V$-$I$_0$=2.12) shows the Li feature. This is better shown in Fig.~6
where we plot the V$_0$ vs. (V$-$I)$_0$ C-M diagram for
the two clusters. Both figures suggest that
the so-called Li chasm (i.e., the drop-off
in lithium detections at the cool temperature end --e.g., Basri \cite{bas97}) 
may
start at earlier spectral--types in IC~2391 than in IC~2602. We
cannot exclude that this is a selection effect, since IC~2602 is better
sampled than IC~2391 at very red colors; in particular, there
are no IC~2391 members in the color interval 1.55$\leq$\vmo$\leq 1.75$ and
only two stars later than \vmo$=2$ are included in our sample
(to be compared with 12 stars in IC~2602). However, under the working
assumption
that low mass stars in the two clusters have the same Li distribution and
that the Li-chasm starts at a similar color (that of R~67), we can
compute the probability of having 0 Li detections among 6 IC~2391 stars in 
the interval \vmo$=1.75-2.15$.
Considering a Li detection rate of 2/7 for IC~2602 and using a
binomial distribution, we find that
the probability of 
having 0/6 detections for IC~2391 is $\sim~13$\%; although this is not
low enough to claim that the two samples are different on a statistical
basis, it does give a hint that the Li distributions of low mass stars
in the two clusters are indeed different. We mention in passing
that neither of the two very late--type stars in IC~2602 showing a Li
detection (R~24A and R~67) have a peculiar position on the C-M diagram
and they well fit into the cluster locus; in other words, they do not seem
to be photometric binaries
(nor were they classified as possible spectroscopic binaries
by S97) or significantly younger than the other cluster members. 
At the same time, all the other very cool stars in IC~2602
do not show any
detectable lithium. Whereas for stars as late as R~67 this might be due
to poor S/N, the EW that we measure for R~24 is much above the
upper limits that we measure for stars of a similar color.
The detection of R~24 and R~67, together 
with the lack of detections for all other stars, witnesses,
therefore, the presence of a spread also among the lowest mass stars
in the IC~2602 sample. We note that the abundances we derived
for R~24A and R~67 (see Table~2) are in any case very low, indicating
a factor of $\sim$ 1000 Li depletion.
\subsubsection{Comparison with the Pleiades}
The left-- and right--hand panels of Fig.~7 
are similar to Figs.~4 and 5, but IC~2602 
and IC~2391 stars cooler than 5300~K are now compared to the Pleiades.
Pleiades EWs from the literature were analyzed consistently
with our sample stars. 
We caveat that, whereas
the two IC~samples are X--ray selected or partially X--ray selected
(in the case of IC~2391), the Pleiades sample is
not, which could introduce some selection effect in the comparison.
This point will be better discussed in Sect.~5.2.

The IC~2602 vs. Pleiades pattern in the left hand panel of Fig.~7 
is generally consistent with the one in Fig.~7 of R97. 
On the one hand this assures that the \nli~vs. \teff~pattern
does not depend
on the particular \teff~vs. color calibration chosen 
while, on the other hand, it
confirms the results of R97 based on a smaller IC~2602 sample.
As mentioned above, the existence of some spread among IC~2602 late--G/early--K 
stars is supported by the present data-set,
although the dispersion among early--K stars
in Figs.~4 and 7 appears somewhat smaller than in
Fig.~6b and 7 of R97, since one of the stars in their sample, R~80,
has been shown to be a cluster non--member by S97. Furthermore,
our present observations 
prove that also later--type stars in the cluster are characterized by
a dispersion in lithium as is the case for the Pleiades. On the contrary,
even a sample of IC~2391 stars much larger than the one of S89 does not
clearly evidence any major spread either among early--K type stars or among
late--K and M dwarfs in the cluster.

Figure~7 also shows that below $\sim$ 4500~K the bulk of
IC~2391 stars are no more lithium rich than the Pleiades.
The \nli~vs. \teff~and EW(Li) vs. (V--I)$_0$ distributions
for IC~2391
lie within the distributions of the Pleiades, with some of the 
IC~2391 stars
close to the Pleiades lower envelope.
As discussed by R97, in the $\sim$ 5600
-- 4000 K range, Pleiades stars at a given temperature 
are less massive than similar temperature stars in the younger IC
clusters.  The other way around,
if we fix the mass, stars at 120 Myr are 
predicted to be warmer
than similar mass stars at 30 -- 50 Myr, since, for masses $\geq$ 
0.6~M$_{\odot}$ (and ages $\ge$ 20~Myr), 
PMS tracks are parallel to the \teff~axis.
Even assuming that no additional Li depletion occurs
between the IC~2391 and the Pleiades ages, the \nli~vs. \teff~ distribution
of IC~2391 between $\sim 4400$ and 3600~K, at an age of 120 Myr,
would appear shifted towards the left
of the diagram by 50--150~K (depending on the star temperature) 
and consequently
the IC~2391 locus would fall on the lower bound of the Pleiades locus.
The situation for IC~2602 is not as clear--cut:
stars warmer than $\sim$ 4400~K and cooler than $\sim$
3900 K in IC~2602 still lie above the mean Pleiades locus, but around
4000~K IC~2602 stars exist that are as depleted as the Pleiades.
Again, if one considers the \nli vs. \teff~distribution that IC~2602
would have at the age of the Pleiades this would be most evident.
\section{Discussion}
\subsection{The cluster ages}
The usually quoted upper main-sequence turnoff age for both
IC~2602 and IC~2391 is $\sim 35$~Myr. S97 carried out a detailed analysis
comparing the observed cluster C-M diagrams with
theoretical isochrones and concluded that the two clusters have a similar
age $\sim 30\pm5$~Myr.
As mentioned in the Introduction,
using the lithium depletion boundary method (LDB --e.g., Stauffer \cite{sta00}
and references therein),
Barrado y Navascu\'es et al. (\cite{bar99}) have derived an age for IC~2391
of 55$\pm$5 Myr; we recall that
an age of 120 Myr
has been inferred for the Pleiades using the same method (Stauffer et al.
\cite{sta98}).
So far no similar study has been carried out for IC~2602 and, therefore,
the question arises whether it is also older than the age inferred
from C-M diagram fitting and whether our present
data allow us to put constraints on its age. 

As discussed by Mart\'\i n (\cite{mar97}), the physics on the cool side of
the so called `` Li chasm" is relatively simple, since the stars are
fully convective; this is not the case on the hot side of the chasm 
and, actually, having  good Li data for a
sample of stars in clusters of different
ages would provide useful feedback to the theory. Standard models predict
that the effective temperature at which lithium is totally depleted 
also
depends on age (e.g., D' Antona \& Mazzitelli \cite{dm94}). In
principle, therefore,
the Li vs. \teff~ pattern shown in Figs.~5,~6, and, 7 and, specifically,
the fact that we do not see the lithium line in IC~2391 stars cooler
than (V $-$ I)$_0 \sim 1.75,$ while in IC~2602 
we detected lithium down to (V $-$ I)$_0 = 2.1$,
could suggest that IC~2602 is younger than IC~2391. 
The conclusion that IC~2391 is older than IC~2602, however, 
would be valid if the observed Li abundances
in both clusters were the result of standard depletion only. 
As a matter of fact, a spread is evident among IC~2602 stars cooler than
4500~K and
Li is not detected in {\it all} very cool members of IC~2602. Moreover,
should one consider the Li vs. color distribution of cool stars
as indicative
of the cluster ages, the conclusion from Figs.~7 would also be that
IC~2391 is at least as old as the Pleiades, which
is not the case.
Given this and the fact, stressed by S97, that ``the distance, reddening, and
ages of the two clusters must be very similar (or differences in one
parameters must be fortuitously compensated for by differences in
another parameter) since there is no discernible difference
in the V vs. \vmo~locus of stars for the two clusters", we believe
that, based on the present data only,
the claim cannot be made that IC~2602 is significantly younger than
IC~2391. Waiting for an age determination of IC~2602 based on the Li
boundary method,
we assume in the following discussion that IC~2602 and IC~2391 are about
co-eval.
\subsection{The spread}
Although the origin of the star-to-star scatter in lithium observed in the
Pleiades and in other young clusters does not have a definitive explanation,
the empirical picture appears so far to be rather well delineated.
Li observations of IC~2602 and \object{IC~4665} (Mart\`\i n \& Montes
\cite{mm97}) suggest that the spread among early K--type stars develops
during the PMS (see above). The dispersion is observed in other
young clusters such as Alpha Per (Balachandran et al. \cite{bal96};
Randich et al. \cite{R98}), \object{Blanco~1} (Jeffries
\& James \cite{jj99}), 
and in clusters intermediate in age between the Pleiades and the
Hyades, such as \object{M~34} (Jones et al \cite{jon97}).
For all clusters for which both lithium and rotational data are
available, a relationship is found between \nli~and the projected
rotational velocity v$\sin i$, in the sense that more rapid rotators
appear to have more lithium than slow rotators (but see King et al. 
\cite{kin00b} 
for an alternative view). Randich et al. (\cite{R98})
more specifically showed that rapid rotators lie on the 
upper bound of the \nli~vs \teff~distribution and are not 
characterized by any major dispersion,
while both Li--rich and Li--poor slow rotators exist, implying that
slow rotators themselves contribute mostly to the dispersion.
In addition, the Li--rotation connection does not hold anymore
for stars cooler than 4500--4400~K (e.g., Garc\`\i a L\`opez et al. 
{\cite{gar94}; Jones et al. \cite{jon96}). 

Before we start any discussion about the possible reasons for the
presence/absence of a scatter in the two IC~clusters, a few words
are warranted about the possible biases due to the sample selection;
we refer to S97 for a more detailed discussion on the completeness
of the two IC~cluster samples, summarizing here their main conclusions.
All but one of the IC~2602 members and part of the IC~2391 members used
in the present paper are X-ray
selected; thus, we cannot {\it a priori}
exclude that our samples are biased 
towards high-activity/high rotation stars. 
The fact that all the IC~2391 members of the
survey of S89 (which was not biased by any means) were detected in the
X--ray survey suggests that the X--ray based membership list for this cluster
should
not be extremely biased. To this respect, note that seven out of the nine
cluster members in the survey of S89 have \vmo~$>0.8$, i.e., are in the
relevant color range for this discussion.
In addition, the comparison with the Pleiades core X--ray surveys
(Stauffer et al. \cite{sta94}; Micela et al. \cite{mic96}) suggests
that the X--ray based membership lists for G and early--K dwarfs 
in the central 50\% of the surveyed
regions of the two IC~clusters should be complete and unbiased.
More specifically, the limiting sensitivity of the IC~2602 X--ray
survey for a region corresponding to about half of the total area
was $\log \rm L_{\rm X} \sim 28.7$~erg/sec. 
Figure~13 in Stauffer et al. (\cite{sta94}) and Fig.~4 in
Micela et al. (\cite{mic96}) clearly show
that virtually all G and early--K Pleiades stars have X--ray luminosities
larger than the above value (inspection of the tables in the two papers
shows that only one star in the range 0.8 $\leq$~\vmo~$\leq$~1.05 has 
a luminosity below that). As argued by S97, this means that,
under the entirely plausible assumption that 
the level of X--ray activity decays with age, at the age of the 
IC~clusters these stars would have had larger X--ray emission and therefore
a survey with a limiting sensitivity $\log \rm L_{\rm X}=28.7$~erg/sec would
be complete. A similar argument holds for IC~2391.}
On the contrary,
members detected in the outer portions of the X--ray images could be
more biased towards active stars/rapid rotators; nevertheless,
as concluded by S97, since for half of the regions
the samples should be rather complete, their velocity (lithium?) range
should be representative of the whole cluster population, unless one
makes the not very likely hypothesis that some sort of velocity
segregation is present within the clusters.
On the other hand, very slow rotating later--type members
may have been missed by the two IC X-ray surveys
and thus a bias may indeed be present for these stars.

In Fig.~8 we plot 
the \nli~vs. \vmo~distributions for the two IC clusters, distinguishing
between slow and rapid rotators. 
Around \vmo~$\sim 0.9-1$, fast rotators in IC~2602 tend to lie on the upper
bound of the distribution, while slow rotators show some dispersion
in Li; the Li--rotation dependence, however, is not as evident as for
the Pleiades or Alpha Per clusters and the sample is too small. 
As is the case for the Pleiades,
any such relationship, if it exists, breaks down below~\vmo $\sim 1.15-1.2$
(or \teff~$\sim 4500$~K).
As to IC~2391, most of the stars in the $0.8 \leq$~\vmo~$\leq 1.15$
color range rotate faster than 15~km/s. 

Low numbers statistics could be the most obvious explanation for the reduced
dispersion observed among stars in IC~2391. In particular,
the relatively small number of late--K IC~2391 stars may certainly explain why
they do not show a significant scatter in Li. Larger and not
activity selected cluster samples are certainly needed to definitively
conclude that the difference in the Li patterns of late--type members in
the two clusters is real. Nevertheless,
as a working hypothesis, we would like to assume
that the difference is indeed real and to make the speculation
that a different distribution in rotational
velocities could instead be the reason for the lack of a major spread among
late--G/early--K stars in IC~2391. More specifically,
if for stars earlier than \vmo$\sim 1.15$
slow rotators are the main contributors 
to the spread, a smaller fraction of slow 
rotators in IC~2391 with respect to IC~2602 (and to the other young 
clusters showing a spread) could explain the different amounts of dispersion;
the fraction of stars in the 0.8~$\leq$ \vmo$\leq 1.15$ interval
with v$\sin i \leq$ 15 km/s in IC~2602 is in fact
larger than in IC~2391 (5/10 against 2/8 --see Fig.~8\footnote{
Under the hypothesis that the velocity distributions
of the two clusters are similar, we computed the probability of
having 2/8 stars with v$\sin i \leq 15$ in IC~2391, given the 5/10 stars
with v$\sin i \leq 15$ in IC~2602. We found that this probability is
equal to 11\%; 
this, again, does not provide a firm proof that the two
distributions are different, but provides a hint that they may be different.}.
Note that the above considerations are based on projected rotational velocities
and we cannot exclude that the IC~2602 sample contains several rapid
rotators seen pole-on. However, Barnes et al. (\cite{barn99} -see
their Fig.~8) suggest that the number of stars with \vmo$\sim$~0.9
with long rotation periods is somewhat larger in IC~2602 than in IC~2391.
To conclude, whereas the weak dependence of Li abundance upon rotation
shown by Fig.~8 and the small number statistics
make this interpretation speculative at this stage,
if the lack of a dispersion in \nli~among IC~2391 early--K type stars
is real and really due to a different distribution in rotational velocities,
the hypothesis can be made
that this in turn may be due to a different fraction
of stars with long-lived disks in the two clusters and, eventually, to 
different 
conditions of the progenitor molecular cloud.
\subsection{Below 4500~K}
Figure~7 supports the results of S89 and R97: IC~2391 low mass
members and part of
IC~2602 very cool stars appear as depleted as the Pleiades.
As a consequence, if Li depletion
does not stop before the stars arrive on the ZAMS, at the Pleiades age they
would be more depleted than the Pleiades. 
In other words, Li depletion in several of the Pleiades cool stars appears to be
slower than in the IC clusters. We stress again that this is not
the case for stars warmer than about 4500~K, implying that such an effect
becomes evident when convection is the major Li destruction mechanism.
As already discussed, we cannot rule out the hypothesis
that the difference between
the Li patterns of the two IC~clusters is due to low number statistics
or possible observational biases. However, we think that
it is unlikely that the lack of a significant larger Li depletion
in the Pleiades with respect to {\it both} IC~clusters is 
also explained by the relatively small sample sizes. 
We also mention that a similar result was found by Randich et al. (\cite{R98})
for the Alpha Per cluster, which is also younger than the Pleiades, but
whose late--type members do not appear more Li rich than the latter.

Under the first order assumption that PMS Li depletion in low mass stars
is driven by convection only, the different depletion rates between
the IC~clusters (and Alpha Per) and the Pleiades can be explained
by a difference in their chemical composition. More precisely,
in order to explain the lower Li depletion, the Pleiades should be
the most metal poor cluster (or have the lowest $\alpha$-element abundances)
among the four.
According to Ventura et al.
(\cite{ven98}) a 15 \% lower than solar metallicity would reduce by a factor 
of 2 PMS Li
depletion of a solar mass star; the effect is obviously larger for lower mass
stars as shown for example by Chaboyer et al. (\cite{cha95} -their Fig.~3).
Our work allows excluding large metallicity differences
between the IC~clusters themselves and between them and
the Pleiades; nevertheless,
given the errors involved, we cannot
exclude differences of the order of 15 \% (or 0.06~dex) in their iron content
nor can we exclude that
the IC clusters have a higher abundance of oxygen or other elements
such as Mg or Si
and thus deplete Li more efficiently on their way to the MS. A more detailed
abundance analysis is clearly needed.

More in general, it is now clear that PMS Li depletion
among cool stars is mainly determined by convective mixing, but it is likely
to depend
on some additional still unknown factor, which either inhibit or accelerate
Li depletion; a difference in this unknown factor would then be the
reason for the spread observed among the Pleiades and IC~2602 stars.
Note that the possibility of 
a major age dispersion among the Pleiades and IC~2602 members
has been excluded by Soderblom et al. (\cite{sod93}) and S97, respectively.
Theoretical models have been constructed showing
that high rotation and the presence of
magnetic fields (which are connected to rotation
via the dynamo mechanism) may inhibit and/or increase
Li depletion. Mart\'\i n
\& Claret (\cite{mc96}) suggested that rapid rotation should inhibit
Li destruction among low mass stars; on the other hand,
Mendes et al. (\cite{men99}) found that fast rotation decreases Li depletion
in fully convective stars, but increases it in presence of
radiative core; their conclusion was that the structural effects of rotation
on PMS Li depletion in low mass stars are not able to bring theory
in agreement with observations. Ventura et al. (\cite{ven98}) and, 
more recently, D' Antona et
al. (\cite{dan00}) have studied the effect of the presence of a magnetic
field on Li depletion; they found\footnote{Calculations are presented for
a 0.95~M$_{\odot}$ only.} that even a relatively small
field, independently on the convection treatment, affects the efficiency
of convection leading to lower Li destruction.
As a matter of fact, however, nor the Pleiades
neither the IC~2602 stars below 4500~K (nothing can be said
about IC~2391 for which we observe no Li dispersion)
show any evident Li--rotation-activity (which is the standard proxy
of surface magnetic fields)
correlation supporting theoretical prediction. 
\section{Summary and conclusions}
We have discussed lithium abundances in the young clusters IC~2602 and
IC~2391 for X-ray selected samples of cool stars whose {\it bona fide}
membership has been confirmed on the basis of 
radial velocity measurements and various spectroscopic indicators. 
We have also determined for the first
time the metallicity of the two clusters which turned out to be
[Fe/H]$=-0.05 \pm 0.05$ and $-0.03 \pm 0.07$, for IC~2602 and IC~2391,
respectively. A reanalysis of two Pleiades stars gives 
[Fe/H]$=-0.03 \pm 0.1$.

The comparison of the Li EWs and Li abundances for IC~2602 and IC~2391 
stars with the Pleiades confirms the previous finding by R97
that late--G and early--K stars in IC~2602 already present a star
to star scatter in Li abundances similar to, although not as large as the
one in the Pleiades. This indicates that the scatter is already present 
at $\sim$35 Myr and must develop during the PMS. IC~2391 seem
to show less
scatter than IC~2602; our present samples do not allow us
to exclude that this is not due to low number statistics
and/or observational biases. If this finding is due to a
real difference in the distributions, it suggests
that the amount of scatter at any given age 
(i.e. the mechanisms responsible for the scatter) may depend on the 
individual cluster and on possibly different initial conditions. 
Whereas the scatter appears to be related to
rotation for early--K Pleiades and the Alpha Persei stars,
we find less evidence for a dependence
on rotation in IC~2602. 

Stars more massive than $\sim 1 M_{\odot}$ show no sign of depletion
in both clusters, while cooler stars are all lithium depleted, with the
amount of depletion increasing to cooler temperatures. The distribution
of Li abundances vs. \teff~ is similar for the two IC clusters down to
$\sim$5000 K, but at lower temperatures the stars of IC~2391 tend to have
less lithium than the stars of IC~2602 of comparable temperature. 
Our data also suggest that the
lithium ``chasm'' may start at bluer colors in IC~2391 than in
IC~2602, although our samples are again too small to claim any statistical
significance for this result. 
 
Some of the coolest stars in IC~2602 (and a fortiori 
in IC~2391) appear in fact 
as depleted as the lowest lithium stars in the Pleiades, 
once proper account is taken of the effective temperature variation, for
a star of a given mass, between the age of the IC~clusters and the Pleiades;
this means that
at the age of the Pleiades, part of the stars in IC~2602 and most
of IC~2391 members are expected 
to be {\it more} lithium depleted than stars of the same mass in the
Pleiades, implying an overall slower Li destruction in the latter cluster.
\section{Appendix}
Several sources of uncertainty affect the final iron abundances;
more specifically, one has to consider (1)
errors which act on individual lines ($\sigma_1$ hereafter and in Table~5)
including random errors
on EW measurements, on gf-values, and on damping constants and  (2) errors
which affect the entire set of lines ($\sigma_2$ hereafter and in Table~5),
the latter mainly including random uncertainties
on atmospheric parameters. To these, possible external errors
originating from the effective temperature and/or microturbulence scale
and from the choice of model atmospheres have to be added.

The standard deviation of the mean abundance from all the analyzed lines
provides a reasonable approximation for $\sigma_1$.
Typical errors in our measurements  of Fe~{\sc i}
line EWs are of the order of 5--15~m\AA. This reflects into a difference
in iron abundance between $\delta \log \epsilon$(Fe)$\sim 0.1$~dex (for
$\delta$~EW=5~m\AA~and a strong line) and
$\delta \log \epsilon$(Fe)$\sim 0.3$~dex (for
$\delta$~EW=15~m\AA~and a weak line) which has to 
be divided by $\sqrt N$, $N$ being the number of lines,
to infer the error in the final Fe abundance for each star.
Correspondingly, considering that we have used between three and seven lines,
errors in the range $\sim 0.04-0.17$~dex are expected.
To these, errors due to uncertainties in atomic parameters have to be added;
in order to estimate them, we changed gf-values and enhancement
factors (E$_{\gamma}$) by 
10 and 50 \%, respectively: the change of gf-values resulted
in a change of the iron abundance (relative to the Sun) of $\sim 0.015$~dex,
while the change in E$_{\gamma}$ resulted in a change of $\sim 0.01-0.02$~dex
(depending on the microturbulence). This, besides showing that errors
in EWs dominate with respect to errors in atomic parameters, indicates
that our assumption for $\sigma_1$ errors is reasonable.

Errors due to uncertainties in stellar parameters (\teff, $\log \rm g$,
and $\xi$) were estimated 
by running the abundance code varying one of the parameters without
changing the others. This was done for the coolest and the warmest stars
in our sample (R~29 and R~21, respectively); 
for each of the three parameters,
the largest variation in \efe~was then taken as a
conservative estimate of the error; more specifically, 
we obtain that 
$+/-$ 100 K
uncertainty in T$_{\rm eff}$, $+/-$ 0.25 dex in $\log\rm g$, and $+/-$ 0.3 km/s
in $\xi$, reflect into errors of 0.072, 0.028, and 0.05  dex in
$\log \epsilon$(Fe), respectively. Assuming errors in surface 
gravities and microturbulence velocities of the order of
0.25~dex and 0.3~km/sec and considering errors in effective temperatures
($\delta \rm T_{\rm eff}$) 
listed in Table~2 (or inferred in the same way for stars in R97 sample and
the two Pleiades stars), the final error in abundance due to
errors in the atmospheric parameters is:\\
$\sigma_2=\sqrt{(\frac{\partial \log \epsilon(\rm Fe)}{\partial \rm T_{\rm eff}}
\times{\delta \rm T_{\rm eff}})^2 + 0.028^2+0.05^2 }$, with 
$\frac{\partial \log \epsilon(\rm Fe)}{\partial \rm T_{\rm eff}}
=\frac{\delta\log\epsilon 
(\rm Fe)_{(\delta \rm T_{\rm eff}=100)}}{100 \rm K}$.

External uncertainties are more difficult to assess.
No trends of iron abundance vs. EW were found, neither a systematic difference
between abundances from low and high excitation lines, suggesting that
the effective temperature and microturbulence scales should not be 
largely in error. On the other hand, whereas
the effective temperature quoted for the two Pleiades stars by King et al.
(\cite{kin00a})
is very similar to the temperature inferred by us, the assumed
microturbulence values
differ by 0.6~km/sec (which would lead to a difference in [Fe/H] 
of the order of $0.07$~dex).
Possible uncertainties due to model atmospheres or, more
in general, to the abundance code were estimated by deriving [Fe/H]
for the Pleiades star hii~676 
using the same atomic and stellar parameters as King et. al.,
but our abundance code. The mean abundance (relative to the Sun)
inferred with our code is [Fe/H]=$-0.011$ to be compared with [Fe/H]$=+0.034$
obtained as the average abundance from the same lines analyzed
by King et al., suggesting
an external error in [Fe/H] of the order of
0.05~dex. 

\begin{acknowledgements}
This research was supported by NSF grants AST-9618335 and AST-9819870 to
SCB. We thank the referee, Dr. R. Jeffries, for his very useful comments
and suggestions.
\end{acknowledgements}

{}
\clearpage
\begin{figure}
\psfig{figure=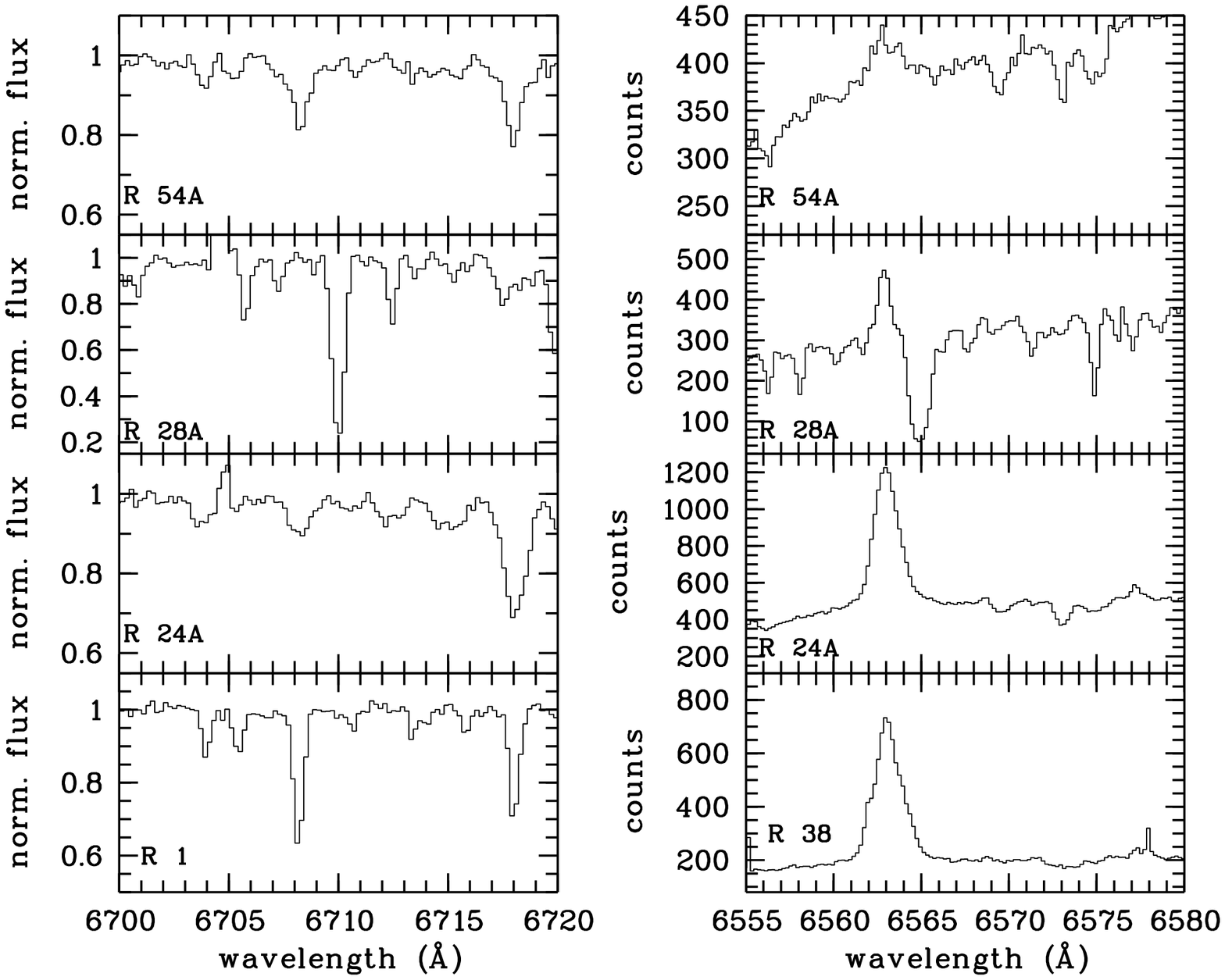, width=12cm}
\caption{Sample spectra around the lithium line (left--hand panels) 
and H$\alpha$ (right--hand panels). The spectra have not been corrected
for radial velocity shifts.}
\end{figure}
\clearpage
\vskip 0.1cm
\begin{figure}
\psfig{figure=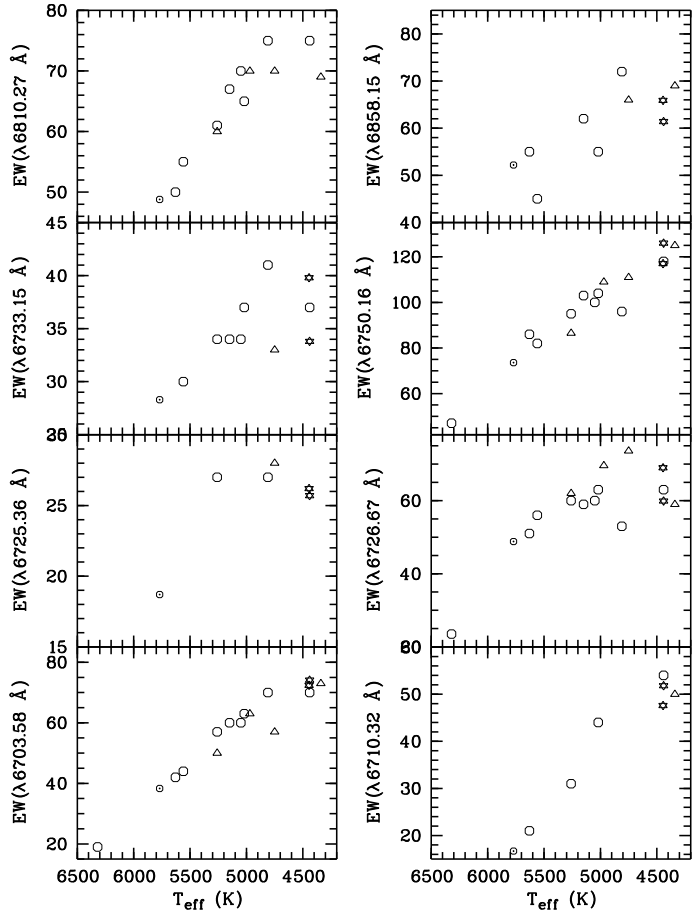, width=12cm}
\caption{EW vs. \teff~for the eight Fe~{\sc i} lines used in our metallicity
analysis for IC~2602 (filled circles), IC~2391 (open triangles),
and the Pleiades (star symbols). The solar EWs also plotted in the figure.}
\end{figure}
\clearpage
\vskip 0.1cm
\begin{figure}
\resizebox{\hsize}{!}{\includegraphics{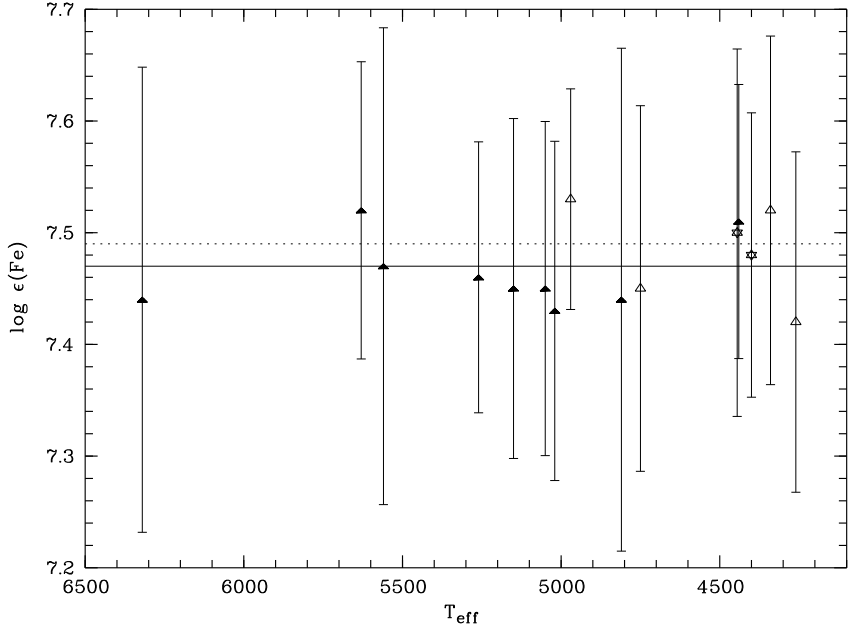}}
\caption{$\log \epsilon$(Fe) vs. effective temperature for the IC~2602
(filled symbols) and IC~2391 (open symbols) stars used to derive metallicities.
Star symbols denote the two Pleiades stars hii~97 and hii~676
analyzed in the same fashion as our
sample stars.
Vertical bars represent the internal errors in iron abundance given by
the sum of $\sigma_1$ and $\sigma_2$ values provided in Table~5.
The horizontal lines denote the weighted average for the three clusters
(solid: IC~2602; dotted: IC~2391 and Pleiades).} 
\end{figure}
\clearpage
\vskip 0.1cm
\begin{figure}
\resizebox{\hsize}{!}{\includegraphics{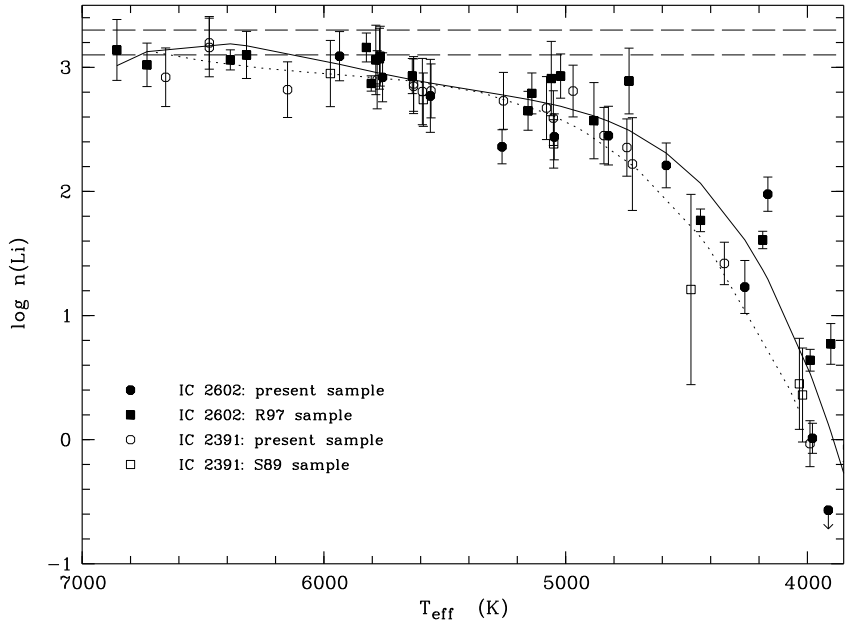}}
\caption{Lithium abundance vs. \teff~for IC~2602 (filled symbols) and IC~2391
(open symbols) members warmer than 3800~K. Circles denote stars belonging
to the present sample, while squares represent stars in R97 and S89 samples.
The two curves are the regression curves of the
$\log$~n(Li) vs. \teff~distributions of the two clusters (solid: IC~2602;
dashed: IC~2391). The long-dashed lines indicate the mean Li abundance
of Classical and Weak lined T Tauri stars and the meteoritic value.}
\end{figure}
\clearpage
\vskip 0.1cm
\begin{figure}
\resizebox{\hsize}{!}{\includegraphics{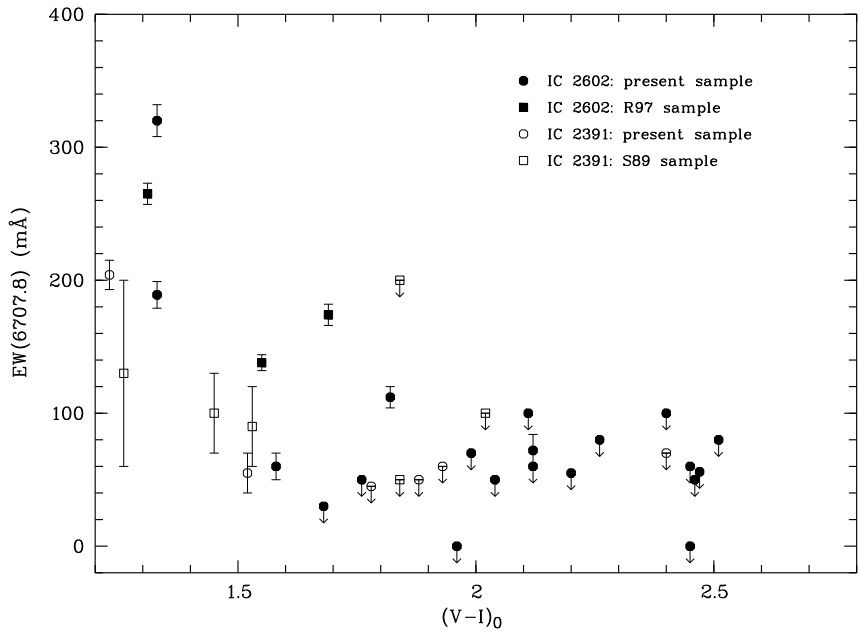}}
\caption{Measured equivalent widths of the Li~I~(6707.8 \AA) line as a 
function of dereddened
(V--I)$_0$~color. Symbols are the same as in Fig.~4.}
\end{figure}
\clearpage
\vskip 0.1cm
\begin{figure}
\resizebox{\hsize}{!}{\includegraphics{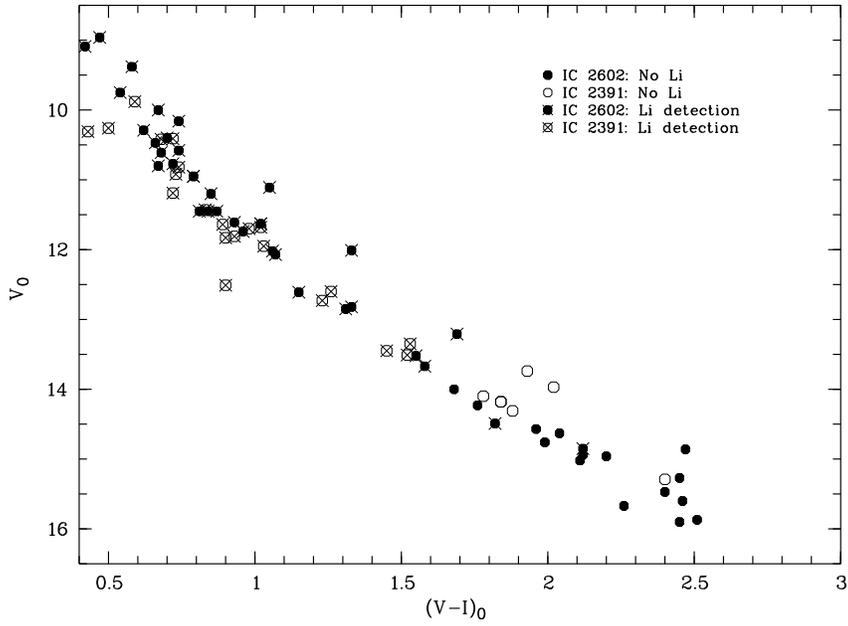}}
\caption{V$_0$ vs. (V--I)$_0$~C-M diagram for IC~2602 (filled
circles)
and IC~2391 (open circles) members. Stars with lithium detection are indicated
by crossed symbols. Note that the two S89 stars SHJM~4 and SHJM~5 have same V 
magnitude and \vmi~color and thus appear as one symbol in the figure.}
\end{figure}
\clearpage
\vskip 0.1cm
\begin{figure}
\psfig{figure=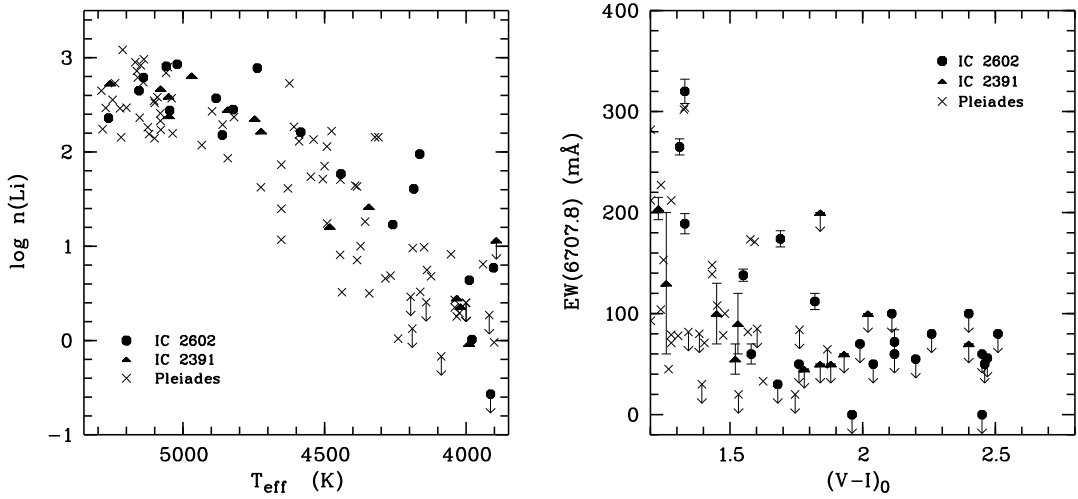, width=12cm}
\caption{Comparison of the \nli~vs. \teff~ distributions of the IC clusters
(filled symbols --circles: IC~2602; triangles: IC~2391) and the Pleiades
(crosses). Only stars cooler than 5300~K are included in the figure.
In the left--hand panel we plot the \nli~vs.\teff~distributions, while
in the right--hand panel the measured Li EWs vs. (V--I)$_0$~are shown.}
\end{figure}
\clearpage
\vskip 0.1cm
\begin{figure}
\resizebox{\hsize}{!}{\includegraphics{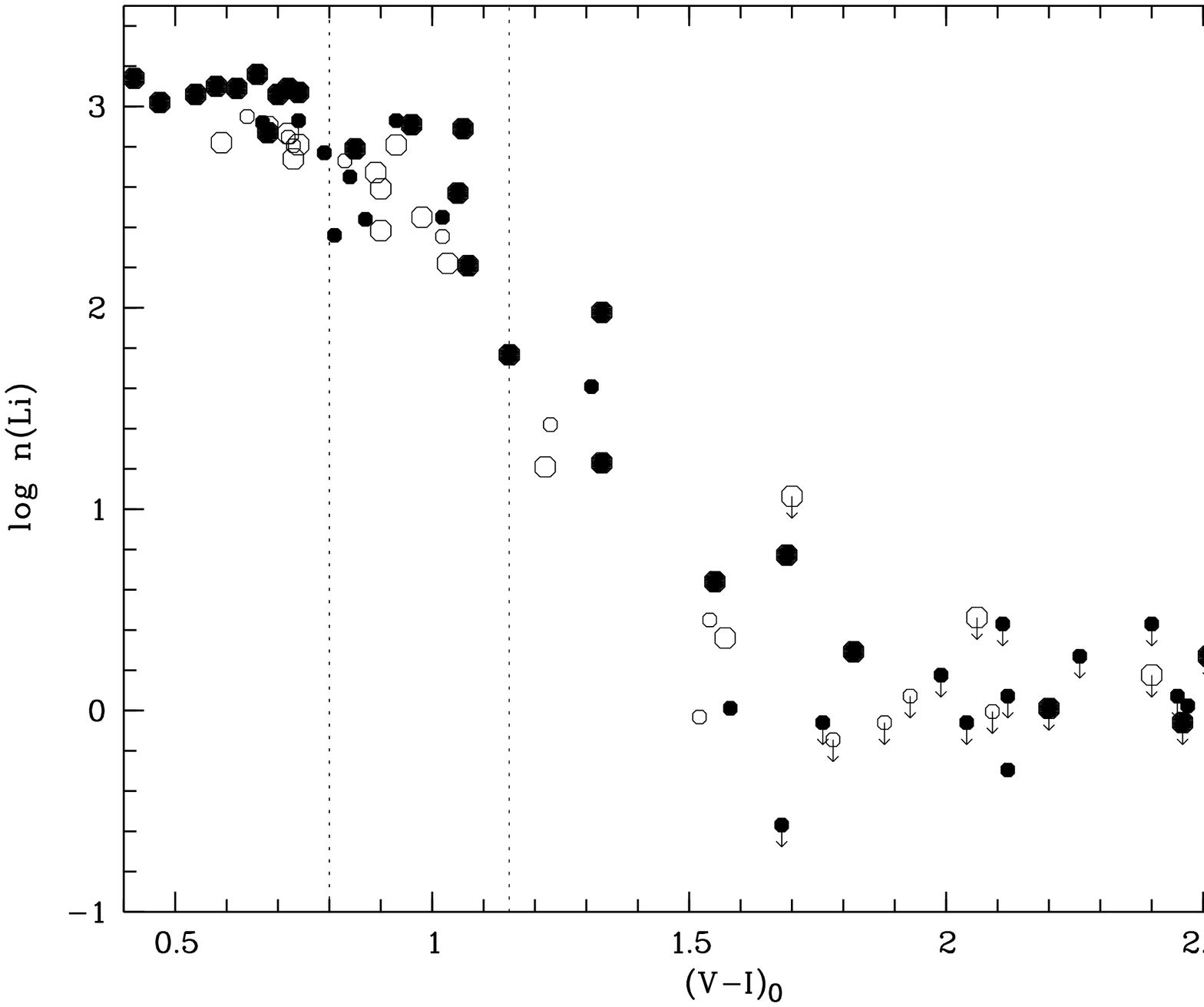}}
\caption{The \nli~vs. \vmo~distributions for the IC~clusters are 
plotted again to investigate a possible dependence on rotation.
Filled and open circles denote IC~2602 and IC~2391 stars, respectively.
Large symbols indicate stars
with v$\sin i > 15$ km/s, while small symbols indicate those with
v$\sin i \leq 15$ km/s.}
\end{figure}


\begin{thebibliography}{}

\bibitem[1996]{bal96}Balachandran, S., Lambert, D.L., 
and Stauffer, J.R., 1996, ApJ, 470, 1243
\bibitem[1999]{barn99}Barnes, S.A., Sofia, S., Prosser, C., 
and Stauffer, J.R., 1999, ApJ, 516, 263
\bibitem[1999]{bar99}Barrado y Navascu\'es D., Stauffer, J.R., 
and Patten, B.M., 1999, ApJ, 522, L53
\bibitem[1997]{bas97}Basri, G., 1997, Mem. SaIt, 68, 917
\bibitem[1979]{bes79}Bessel, M.S., 1979, PASP, 91, 589
\bibitem[1991]{bes91}Bessel, M.S., 1991, AJ, 101, 662
\bibitem[1987]{bw87}Bessel, M.S., and Weis, E.W., 1987, PASP, 99, 642
\bibitem[1990]{bf90}Boesgaard, A.M., and Friel, E.D., 1990, ApJ, 351, 467
\bibitem[1962]{bra62}Braes, L.L.E., 1962, Bull. Astron. Inst. Neth., 16, 297
\bibitem[1994]{car94}Carlsson, M., Rutten, R.J., Bruls, J.H.M.J., 
and Shchukina, N.G., 1994, A\&A, 288, 860
\bibitem[1997]{cg97}Carretta, E., and Gratton, R.G., 1997, A\&AS, 121, 95
\bibitem[1995]{cha95}Chaboyer, B., Demarque, P., and Pinsonneault, M.H., 1995,
ApJ, 441, 876
\bibitem[2000]{dan00}D' Antona, F., Ventura, P., and Mazzitelli, I., 
2000, ApJ, 543, L77
\bibitem[1994]{dm94}D' Antona, F., and Mazzitelli, I., 1994, ApJS, 90, 467
\bibitem[2000]{del00}Deliyannis, C.P., 2000, in
Stellar Clusters and Associations: Convection, Rotation, and Dynamos,
R. Pallavicini, G. Micela, and S. Sciortino (eds), ASP Conf. Ser. 198, 235
\bibitem[1994]{gar94}Garc\'\i a L\'opez, R.J., Rebolo, R., 
and Mart\'\i n, E.L., 1994, A\&A, 282, 518
\bibitem[1991]{gs91}Gratton, R.G., and Sneden, C., 1991, A\&A, 241, 501
\bibitem[2000]{jef00}Jeffries, R.J., 2000,  in
Stellar Clusters and Associations: Convection, Rotation, and Dynamos.
R. Pallavicini, G. Micela, and S. Sciortino (eds), ASP Conf. Ser. 198,
245
\bibitem[1999]{jj99}Jeffries, R.D., and James, D.J., 1999, ApJ, 511, 218
\bibitem[1998]{jef98}Jeffries, R.D., James, D.J., and Thurston, M.R., 
1998, MNRAS, 300, 550
\bibitem[1996]{jon96}Jones, B.F., Shetrone, M., 
Fisher, D., and Soderblom, D.R., 1996, AJ, 112, 186
\bibitem[1997]{jon97}Jones, Fischer, D., B.F., Shetrone, M., 
and Soderblom, D.R., 1997, AJ, 114, 352
\bibitem[1999]{jon99}Jones, B.F., Fisher, D., and Soderblom, 
D.R., 1999, AJ, 117, 330 
\bibitem[2000a]{kin00a}King, J.R., Soderblom, D.R., 
Fisher, D., and Jones, B.F., 2000,
ApJ, 533, 944
\bibitem[2000b]{kin00b}King, J.R., Krishnamurthi, A., 
and Pinsonneault, M.H., 2000, AJ, 119, 859
\bibitem[1993]{kir93}Kirkpatrick, J.D., Kelly, D.M., 
Rieke, G.H., and Liebert, J., 1993, ApJ, 402, 643
\bibitem[1995]{kur95}Kurucz, R.L., 1995, private communication
\bibitem[1996]{leg96}Leggett, S.K., Allard, F., Berriman, 
G., Dahn, C.C., and Hauschildt, P.H., 1996, ApJS, 104, 117
\bibitem[1997]{mar97}Mart\'\i n, E.L., 1997, Mem. SaIt, 68, 905
\bibitem[1996]{mc96}Mart\'\i n, E.L., and Claret, A., 1996, A\&A, 306, 408
\bibitem[1997]{mm97}Mart\'\i n, E.L., and Montes, D., 1997, A\&A, 318, 805
\bibitem[1994]{mar94}Mart\'\i n, E.L., Rebolo, R., Magazz\'u, A., and Pavlenko,
Ya.V., 1994, A\&A, 282, 503
\bibitem[1999]{men99}Mendes, L.T.S., D' Antona, F., 
and Mazzitelli, I., 1999, A\&A, 341, 174
\bibitem[2000]{meo00}Meola, G., Pallavicini, R., 
Randich, S., Stauffer, J.R., and
Balachandran, S.C., 2000, in
Stellar Clusters and Associations: Convection, Rotation, and Dynamos,
R. Pallavicini, G. Micela, and S. Sciortino (eds), ASP Conf. Ser. 198, 285
\bibitem[1996]{mic96}Micela, G.Sciortino, S.; Kashyap, V.;
 Harnden, F. R., Jr.; Rosner, R., 1996, ApJS, 102, 75
\bibitem[2000]{pas00}Pasquini, L., 2000, in IAU Symposium
198: ``The Light Elements and their Evolution", L. da Silva, M. Spite,
and J.R. de Medeiros (eds), p.~245
\bibitem[1996]{ps96}Patten, B.M., and Simon, T., 1996, ApJS, 106, 489
\bibitem[1995]{pav95}Pavlenko, Y.V., Rebolo, R., 
Mart\'\i n, E.L., Garc\'\i a L\'opez, R.J., 1995, A\&A, 303, 807
\bibitem[1998]{pin98}Pinsonneault, M.S., Stauffer, J.R., 
Soderblom, D.R., King, J.R., and Hanson, R.B., 1998, ApJ, 504, 170
\bibitem[1996]{pro96}Prosser, C.F., Randich, S., 
and Stauffer, J.R., 1996, AJ, 112, 649
\bibitem[1995]{R95}Randich, S., Schmitt, J.H.M.M., Prosser, C.F., and Stauffer,
J.R., 1995, A\&A, 300, 134
\bibitem[1997]{R97}Randich, S., Aharpour, N., Pallavicini, R., Prosser, C.F.,
and Stauffer, J.R., 1997, A\&A, 323, 86
\bibitem[1998]{R98}Randich, S., Mart\'\i n, E.L., Garc\'\i a L\'opez, R.,
and Pallavicini, R., 1998, A\&A, 333, 591
\bibitem[1993]{sod93}Soderblom, D.R., Jones, B.F., Balachandran, S., et al.,
1993, AJ, 106, 1059
\bibitem[1999]{sod99}Soderblom, D.R., King, J.R., Siess, L., Jones, B., 
and Fisher, D., 1999, AJ, 118, 1301
\bibitem[2000]{sta00}Stauffer, J.R. 2000, in
Stellar Clusters and Associations: Convection, Rotation, and Dynamos,
R. Pallavicini, G. Micela, and S. Sciortino (eds), ASP Conf. Ser. 198, 255
\bibitem[1989]{sta89}Stauffer, J.R., Hartmann, L.W., 
Jones, B.F., and McNamara, B.R., 1989, ApJ, 342, 285
\bibitem[1994]{sta94}Stauffer, J. R.; Caillault, J.-P.; Gagne, M.;
 Prosser, C. F.; Hartmann, L. W., 1994, ApJS, 91, 625
\bibitem[1997]{sta97}Stauffer, J.R., Hartmann, L.W., 
Prosser, C.F., Randich, S., Balachandran, S., et al., 1997, ApJ, 479, 776
\bibitem[1998]{sta98}Stauffer, J.R., Schultz, G., and 
Kirkpatrick, J.D., 1998, ApJ, 499, L199
\bibitem[1955]{uns55}
Uns\"old, A., 1955, Physik der Sternatmosph\"aren, Springer-Verlag, Berlin
\bibitem[1998]{ven98}Ventura, P., Zeppieri, A., 
Mazzitelli, I., and D' Antona, F., 1998, A\&A, 331, 1011
\bibitem[1961]{whi61}Whiteoak, J.B., 1961, MNRAS, 123, 245
\end{thebibliography}
\end{document}